\documentclass[twocolumn]{aastex63}
\usepackage{mathrsfs}  
\usepackage{xcolor}
 
\usepackage{graphicx}
\usepackage{amsmath}
\usepackage{lineno}
\usepackage{wasysym}

\newcommand\SystemAnalysed{64 }
\newcommand\PlanetsAnalysed{218 }
\newcommand\SignificantMass{88 }  
\newcommand\LitSignificantMass{73 }
\newcommand\NewSignificantMass{23 }
\newcommand\RelativeMassError{$\sim22\%$ }
\newcommand\SmallPlanets{14 } 
\newcommand\NewSmallPlanets{5 } 
\shorttitle{Planetary Mass Determinations from a Simplified Photodynamical Model}
\shortauthors{Ofir et al.}
\graphicspath{{./}{figures/}}

\begin{document}

\title{Planetary Mass Determinations from a Simplified Photodynamical Model - Application To The Complete Kepler Dataset}

\correspondingauthor{Aviv Ofir}
\email{avivofir@weizmann.ac.il}

\author[0000-0002-9152-5042]{Aviv Ofir}
\affiliation{Department of Earth and Planetary Sciences, Weizmann Institute of Science, Rehovot, 76100, Israel}

\author[0000-0002-1451-6492]{Gideon Yoffe}
\affiliation{Department of Earth and Planetary Sciences, Weizmann Institute of Science, Rehovot, 76100, Israel}
\affiliation{Department of Statistics and Data Science, The Hebrew University of Jerusalem, Jerusalem, 91905, Israel}
\author[0000-0001-9930-2495]{Oded Aharonson}
\affiliation{Department of Earth and Planetary Sciences, Weizmann Institute of Science, Rehovot, 76100, Israel}
\affiliation{Planetary Science Institute, Tucson, AZ, 85719, USA}



\begin{abstract}

We use \texttt{PyDynamicaLC}, a model using the least number of- and the least correlated- degrees of freedom needed to derive a photodynamical model, to describe some of the smallest- and lowest TTV (transit timing variations) amplitude- of the \textit{Kepler} planets. We successfully analyze \SystemAnalysed systems containing \PlanetsAnalysed planets, for \SignificantMass of which we were able to determine significant masses (to better than $3\sigma$). We demonstrate consistency with literature results over two orders of magnitude in mass, and for the planets that already had literature mass estimations, we were able to reduce the relative mass error by \RelativeMassError (median value). Of the planets with determined masses \NewSignificantMass are new mass determinations with no previous significant literature value, including a planet smaller and lighter than Earth (KOI-1977.02 / Kepler-345 b). 

These results demonstrate the power of photodynamical modeling with the appropriately chosen degrees of freedom. This will become increasingly more important as smaller planets are detected, especially as the \textit{TESS} mission gathers ever longer-baseline light curves and for the analysis of the future \textit{PLATO} mission data.

\end{abstract}



\section{Introduction}

Exoplanetary masses may be determined from Transit Timing Variations (hereafter TTVs) due to the mutual gravitational perturbations among the members of a system \citep{Holman05, Agol05}. TTVs are extremely common: they exist in all systems that have more than one planet, and only our ability to detect and interpret them limits this technique's applicability. We previously introduced \texttt{PyDynamicaLC}, a new algorithm \citep{Yoffe21} for the determination of planetary masses from photometric datasets, leveraging the TTVs that occur in multi-planet systems due to their non-Keplerian motion. The algorithm is suitable for a subset of multi-planetary systems, those characterized by low eccentricity, low relative inclinations, and near, but not locked in, first-order mean motion resonance (MMR). This seemingly restrictive subset is relatively common in nature. Such configurations allow us to use several simplifications without loss of accuracy and, importantly, increase the sensitivity of the technique as a whole to ever-lower masses, including measuring the masses of planets not previously constrained. In the first paper \citep{Yoffe21}, we introduced and described the technique, as well as analyzed three Kepler systems with \texttt{PyDynamicaLC}.

Here, we apply \texttt{PyDynamicaLC} to a much larger population using Kepler's long cadence data and report the resulting masses of multiple exoplanets. In \S\ref{SampleAndProcessing} we describe the sample and the processing leading to the TTV analysis, in \S\ref{PhotodynamicalModel} we update the interpretation of \texttt{PyDynamicaLC}'s results relative to  \citet{Yoffe21}, based on additional experience gained in using it. In \S\ref{Results}, we provide the results and conclude in \S\ref{conclusions}.

\section{Sample and Light Curve Processing}
\label{SampleAndProcessing}

Our initial sample is the 440 KOIs from 373 systems that have high confidence TTV ($99\%$) based on bootstrapping analysis, as presented in \citet[Table 1]{Ofir18}. Of these, 163 multi-transiting systems were selected for this study since we aim to model the interaction among known planets. TTVs in singly transiting systems are likely caused by unseen companions, not considered here. Several systems were manually removed due to difficulties in the pre-processing stage (explained below), for example, due to orbiting highly active stars with background variability which is difficult to remove or variable apparent depth due to variable photometric contamination in the aperture in different Kepler quarters. Such transit signals would have residuals from a simple \citet{MA02} model that can be misinterpreted as TTVs. We, therefore, exclude systems with a poor linear ephemeris model - making a final sample of \PlanetsAnalysed systems. We note that in this sample, we do not analyze high-amplitude TTVs (amplitudes comparable to the transit duration or longer, as defined in \citet{Ofir18}), as \texttt{PyDynamicaLC} aims to improve the available measurements in information-poor systems such as low TTV-amplitude or shallow transits systems.

The final Kepler data release DR25 data was obtained from the NASA NExScI database \cite{koidr25} \footnote{\texttt{https://exoplanetarchive.ipac.caltech.edu/index.html}}. The initial data processing included two conventional iterative stages: a detrending/cleaning stage and an empirical model stage. Kepler's long cadence data was first filtered using a cosine filter and then modeled empirically using the Mandel Agol model \citep{MA02} (detailed below), the resultant model of which was then divided out from the raw data to allow better re-filtering in the next iteration, repeating until convergence is reached. Specifically, if a given system includes $N_{\rm pl}$ planets, the filtering process was performed $N_{\rm pl}$ times - with the cosine filter's maximum frequency, $f_{\rm max}$, set such that $f_{\rm max}^{-1}$ was smaller than three times the transit duration of each of the $N_{\rm pl}$ planets. In each continuous section of the light curve (between large gaps of more than 10 data points), we chose the filter corresponding to the longest-duration transit appearing in that section of the data only, thereby allowing stronger filtering when possible without modifying the underlying transit signals. Planets are assumed to have strictly periodic orbits unless they appear in \citet{Holczer16}, in which case the initial guess for the times of mid-transit is taken from there. Although \citet{Holczer16} was a good starting point, we found that a significant number of additions and modifications were needed. Many of these modifications were expected (e.g., \citet{Holczer16} intentionally did not time partial transit or transits close to data gaps) - but some other timings were missing or significantly off, and thus initial guesses that were outliers to the general trend were manually corrected. Occasionally, the modifications were to either add or remove a planet candidate since some signals were found to be false positives by the NExScI database, and some were identified after \citet{Holczer16} (see \S\ref{Results} for system-by-system discussion). The fitting was performed using an implementation of Differential Evolution Monte Carlo \citep{Braak06} and using the Gelman-Rubin convergence criterion \citep{GelmanRubin92}, slightly modified such that if a single chain caused the Gelman-Rubin criterion to fail -- the MCMC was still considered to have converged. We note that the number of chains was chosen to be $N_{\rm DOF}\log\left(N_{\rm DOF}\right)$, and the number of degrees of freedom $N_{\rm DOF}$ is large, corresponding to the individual times of mid-transit in the entire system. Therefore, this modification of the stopping criterion served only to avoid continuing a long MCMC run failing to converge due to a single chain trapped in a local minimum within this high dimensionality space.

Importantly, the goal of the pre-processing stage was not to re-measure the times of mid-transit, which are not used in fitting the global model. The only results progressing to the photodynamical model are the de-trended light curve, and the usual Mandel-Agol parameters of mean ephemeris $P, T_{\rm mid}$ and geometric properties of the transits $a$, $b$, and $r$ for each planet (the normalized semi-major axis, impact parameter, and radius - all normalized to the star's radius). Notably, only the semi-major axis of the innermost planet was fitted, while the rest were scaled according to Kepler's second law (as in \cite{Ofir2013}). We compared the resultant Mandel-Agol parameters of the above procedure with the ones derived by NExSci and found near-complete agreement, with notable disagreement only for grazing planets - where our derived $a,b$ and $r$ are much better constrained (e.g., for KOI-1102.03 $a_\textrm{NExSci}=29.6\pm3.6$ while $a_\textrm{this\,work}=29.56^{+0.39}_{-0.38}$). This difference is expected: grazing planets can only weakly constrain all of $a, b$, and $r$ simultaneously on their own, and only certain combinations of these three parameters are meaningfully constrained. However, due to our procedure's single $a$ common to all planets in a given system, we more strongly constrain $a$ even for grazing planets. This, in turn, allows for better determinations of also $r$ and $b$ (e.g. for KOI-1102.03 above $r_\textrm{NExSci}=0.3^{+73.1}_{-6.9}$ with a similarly very large range for $b$, whereas $r_\textrm{this\,work}=0.02055^{+0.00043}_{-0.00044}$). In the following analysis, we chose to use the NExSci Mandel-Agol parameters for all but the grazing planets (these were defined as those with NExSci fit in which $r+b>1$), and the parameters derived using the above procedure for the grazing planets. The values used for each planet are given in Table \ref{PlanetGeometricParams}.

\section{Photodynamical Analysis} 
\label{PhotodynamicalModel}

In \citet{Yoffe21}, we presented \texttt{PyDynamicaLC} - a simplified photodynamical model tailored to modeling low-eccentricity, zero relative inclination systems to derive planetary masses. It is based on \texttt{TTVFaster} \citep{TTVFaster}, an analytic series expansion of the TTV signal that is accurate to first order in the planet–star mass ratios and in the orbital eccentricities. A nonlinear optimization of the dynamical parameters of $m$, $\Delta e_x$, and $\Delta e_y$ - the mass and eccentricity-difference vector components, and estimates of their uncertainties are performed using \texttt{MultiNest} \citep{Multinest}.  The eccentricity components are $e_x=e\cos\omega$ and $e_y=e\sin\omega$, where $e$ and $\omega$ are the orbital eccentricity and argument of periapsis, respectively, and $\Delta$ refers to the difference between the eccentricity component of a planet and that of the planet just interior to it. For the innermost planet, the values are the eccentricity components themselves.   We use here the same code for the analysis as \citet[][]{Yoffe21} with a minor correction, wherein the successive eccentricity-difference components in high-multiplicity systems were not chained correctly. We also update the post-MCMC validation process, as detailed below. As before, we run sets of simulations with different priors and report preferred solutions within the validity domain of \texttt{TTVFaster}. We found the Validity Map used in \citet{Yoffe21} can be an overly conservative test in verifying solutions. Valid solutions may be rejected due to the inability to translate from mean orbital elements in \texttt{TTVFaster} to instantaneous elements required as initial conditions for the N-body simulation. This is because the Validity Map is a 2D cut through a $3N_{\rm pl}$-dimensional space, which may not pass through the true system parameters. Hence, deviations from N-body integrations along this plane of solutions may overestimate the deviations from the true system.

Here, we ran a set of six MCMC optimizations for each system - all combinations of three eccentricity components samplings and two mass samplings, expanding on \citet{HL17}, which used two optimizations per system. $\Delta e_x$ and $\Delta e_y$ were sampled using three Gaussian distributions ($N(0,0.01), N(0,0.02)$ or $N(0,0.05)$), and the planetary masses were sampled using either a uniform of log-uniform distribution. To overcome the shortcomings of the Validity Map of \cite{Yoffe21}, our approach for selecting the best simulation of these six relies more heavily on the mathematical basis of \texttt{TTVFaster} than heuristics. It consists of several tests that each acceptable solution must pass, followed by the identification of statistically equivalent solutions among the surviving ones. These tests are as follows.

1. Small high-order contribution: 
We use a TTV model utilizing a first-order series expansion. We, therefore, need to ensure higher-order terms do not contaminate our results. We wish to estimate the TTV amplitude of all pairs of planets in the system for several orders of their interaction and require that the first-order estimated TTV amplitude dominates. Because detailed modeling of the TTV signal of high-order near-MMR signal requires all orbital elements of all planets, which are unknown, we use analytical TTV amplitudes based on the first-order estimates in Eqs.~8,9 of \citet{LithXieWu12}, where each planet is assigned an estimated mass $m/m_\oplus=(r/r_\oplus)^{2.5}$, each adjacent planet pair was assumed to have an eccentricity difference of $Z_{free}=0.05$, and higher order interactions of orders $O=[2...4]$ were modeled by multiplying the result by $Z_{free}^{O-1}$. These choices are simple and representative of the multi-planet systems we wish to model. Importantly, these estimates most critically depend on the normalized distance from resonance:

\begin{equation}
    \Delta \equiv \frac{P'}{P}\frac{j-O}{j}-1
\end{equation}

Where $P'$ and $P$ are the outer and inner periods, respectively, of the pair under consideration, and $j/(j-O)$ is their integer-approximated ratio. The normalized distance from resonance is well-constrained from observations and typically leads to a clear distinction between cases where higher-order near-MMRs are significant to those where they are not with a prescription that is insensitive to the precise choices above.  We require that the estimated amplitude for the first-order interaction TTVs be larger than the quadrature sum of all higher-order TTV amplitude estimates.

2. Low eccentricity: \texttt{TTVFaster} is limited to low eccentricities \citep{TTVFaster}. We therefore check that none of the $\Delta e_x$ or $\Delta e_y$ components have a magnitude $>$ 0.1. The results are not sensitive to the precise value chosen. 

3. Out of MMR: \texttt{TTVFaster} is valid close to, but not in, mean motion resonances \citep{TTVFaster}.  We use the analytical formula in Ch.~8 of \cite{Murray2000} to estimate the resonance width of first-order resonances (their eq. 8.74). Because this resonance width is approximate, we require that the observed distance from resonance is at least $1.5$ times the estimated width.

4. Unreasonable densities: similarly to \citet{Judkovsky22b}, we require that the resultant planetary densities will be within the physically plausible range.  Specifically, they are no more than $2\sigma$ higher than a density of $12 g\, cm^{-3}$ - which is close to the theoretical maximum density of an iron planet. On the lower end, the log-uniform mass sampling does not have a well-defined natural lower bound. We therefore require that the derived masses from these solutions are at least $2\sigma$ above the lower boundary of our prior of $0.01 M_\oplus$.

Finally, and similarly to \citet{Judkovsky22b}, we group the surviving solutions by using the method by \citet{Pastore2019}: two solutions are considered similar if all parameters have overlapping PDFs in all parameters. This is the common outcome: most solutions, even those with different priors, converge to a single or a small number of minima in parameter space, and each group of such solutions is represented by its lowest-$\chi^2$ member (which, by definition, is statistically indistinguishable from the other members of the group). We quote multiple solutions whenever the grouping is incomplete and sufficiently different groups still exist and survive the other tests. In practice, none of the systems converged to more than two distinct solutions.

The dynamical model above uses only the relative masses between the planets and the host star. Physical parameters were derived using the stellar parameters of \citet{Fulton2018}, with some radii taken from \cite{Berger18}, and in the few cases where those were not available - from the NExScI database. Table \ref{StellarParams} gives the stellar parameter used in the analysis.

\section{Results} 
\label{Results}

\subsection{Summary statistics}

\begin{figure}
  \includegraphics[width=\linewidth]{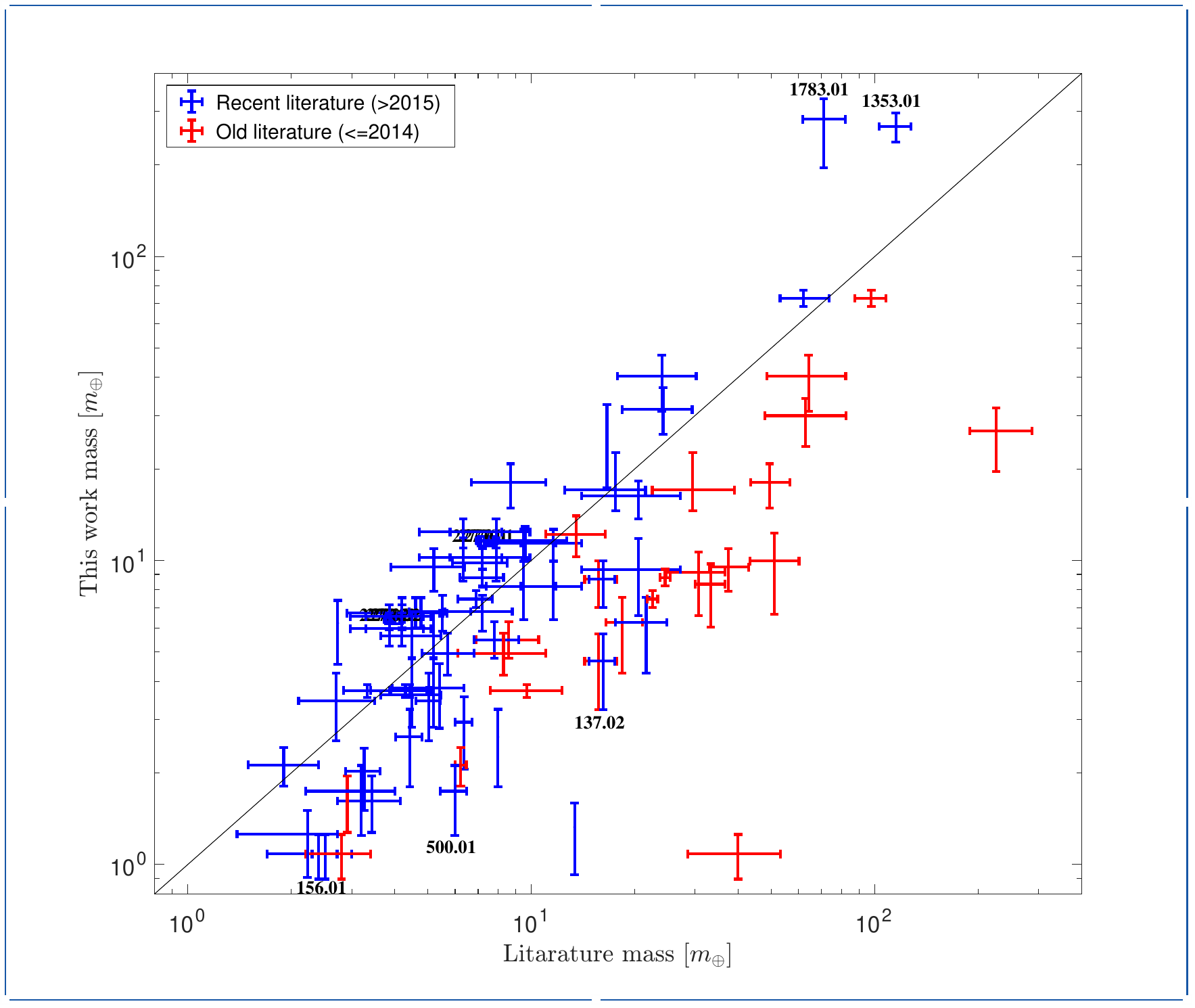}
  \caption{Comparison of the mass estimates of objects with both literature and this work's mass estimates significant to better than $3\sigma$. The red symbols use older references (2014 and earlier), while the blue symbols use newer references (2015 and later). The recent estimates are more consistent with our own than earlier ones - demonstrating increasing agreement between different analyses of the same objects. We labeled systems in which the newer literature mass estimates differ from our own by more than $4\sigma$.
  Note that there are duplicate entries: if multiple solutions are given in the paper - both appear here (with a common abscissa value). Also, if more than one literature source exists - all will be given (with a common ordinate value).}
  \label{MassComparison}
\end{figure}

\begin{figure}
\noindent\includegraphics[width=\linewidth]{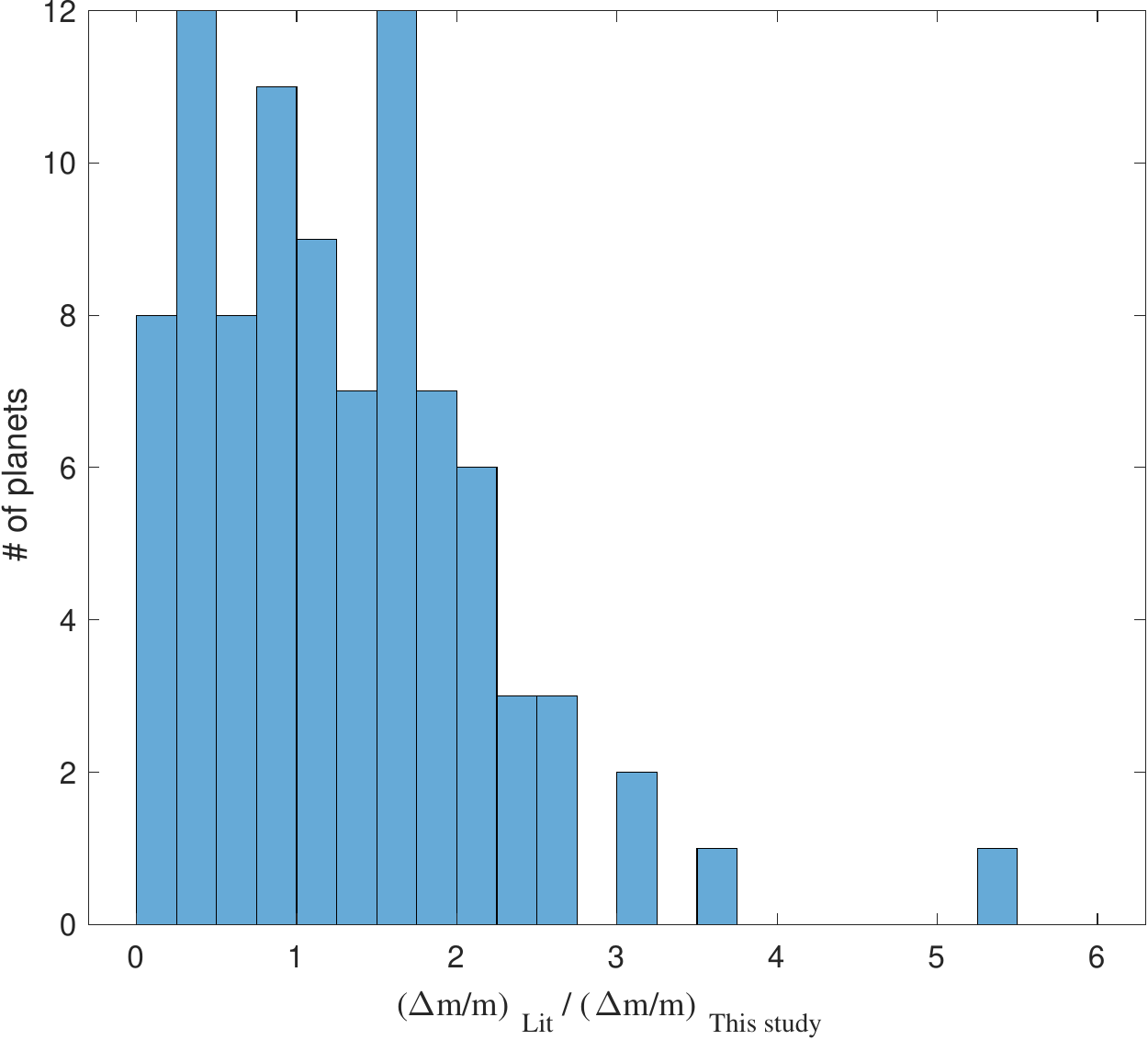}
\caption{\footnotesize Relative mass error comparison for planets with significant masses both in this work as well as in the literature (the ones appearing in Fig. \ref{MassComparison}). The literature median relative mass uncertainty is observed to be about \RelativeMassError higher than the current study's relative mass uncertainty.}
  \label{RelativeErrorComp}
\end{figure}

Dynamical analysis was completed successfully (passing all criteria) on \SystemAnalysed systems containing \PlanetsAnalysed planets, all shown in Table \ref{PlanetParams}. Of these, \LitSignificantMass had prior literature value (which we often improve here). \SignificantMass planets were found to have better than $3\sigma$ mass detections in valid solutions, of them \NewSignificantMass are completely new mass determinations. These new masses are mostly below $10 m_\oplus$
and reach down to less than one Earth mass and one Earth radius (KOI-1977.02 / Kepler-345 b) -- meeting our goal of constraining the masses of low-mass planets. When comparing the masses of planets with significant literature values (see Fig.~\ref{MassComparison}) we find general agreement over two orders of magnitude. Moreover, when we compare the uncertainties of these common mass determinations, we find that the relative mass error ${\Delta m}/{m}$ is at least \RelativeMassError larger (median value) in the literature values than the one given here (see Fig.~\ref{RelativeErrorComp}). The overall agreement in the magnitude of the errors provides increased confidence in our uncertainty estimation, with an appreciable reduction in uncertainty demonstrating the advantage of using the global model and optimized MCMC parameters even when individual times of mid-transit are available \citep{Judkovsky23}.

\begin{figure}
\noindent\includegraphics[trim={1cm 0.5cm 5.2cm 1cm}, width=0.75\linewidth]{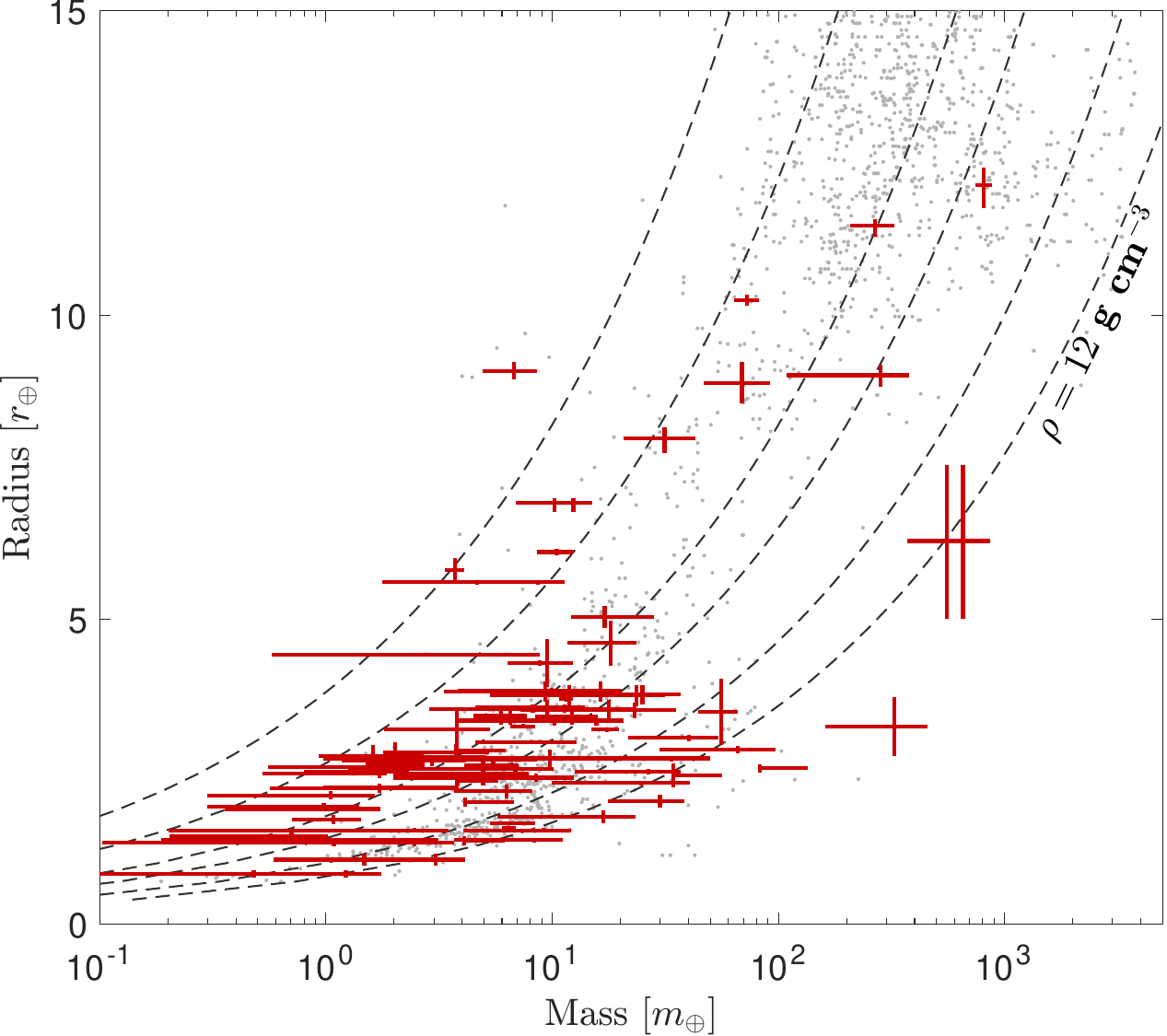}
\caption{\footnotesize The mass-radius plane for all planets with significant mass in this study (maroon error bars). Note that a planet may appear more than once in the plot if more than one group of solutions is arrived at, and these points will share a common radius. Over-plotted are lines of constant bulk density, as well as all planets with mass and radii determined to be better than $3\sigma$ from the NASA NExSci database (light gray points).}
  \label{MassRadiusFigure}
\end{figure}

Fig. \ref{MassRadiusFigure} shows the distribution of planets
measured in this study in the mass-radius plane. The figure also plots previously measured masses from the NEXSci catalog for comparison. The relative lack of massive planets in this study is expected from our target population: firstly, we model only multi-planet systems (where giant planets are less common), and secondly, this study includes only systems that exhibit low-amplitude TTVs as defined by \cite{Ofir18} - and massive planets tend to induce large-amplitude TTVs on their neighboring planets. There is an approximate agreement between the distribution of literature planets and the distribution of planets in this study, but upon closer inspection, one can see a subtle shift: for a given radius, most masses arrived at in this study are lower than the literature masses. Again, this is expected since this study's focus was to extract dynamical information from the lowest TTV amplitude planets, enhancing the sensitivity to lower-mass planets relative to the literature \cite{Judkovsky23}.

Below we report our system-by-system results, in two categories: systems in which at least one new mass determination was made (i.e., a better than $3\sigma$ mass detection where none was known before), and mass determinations for planets with previous literature mass determination. The systems are sorted by KOI number and in planet-star distance order within each system.

\subsection{Individual Systems - New Mass Determinations}

\subsubsection{KOI-94 / Kepler-89}
\citet{Weiss13} used RV to measure the mass of the third (and the largest) of this four-planet system and found KOI-94.01 to have a mass of $m_{94.01}=102\pm11.4 m_\oplus$, while for the other planets significant mass detections were not achieved. These measurements assumed a host mass of $m_\star=1.277\pm0.050 m_\odot$ whereas later analyses corrected that value to $m_\star=1.177^{+0.032}_{-0.024} m_\odot$ \citep{Fulton2018}. Correcting for the lower stellar mass, the planet's mass is $5.5\%$ lower, or $m_{94.01}\approx100\pm11 m_\oplus$. \citet{masuda13} analyzed the TTVs in the system, including the above RVs (and also the identification of a single planet-planet eclipse candidate event). Taken together, they concluded that
$m_{94.02} = 9.4^{+2.4}_{-2.1}m_\oplus$, $m_{94.01} = 52.1^{+6.9}_{-7.1}m_\oplus$, $m_{94.03} = 13.0^{+2.5}_{-2.1}m_\oplus$ -- meaning that there is a $3-4\sigma$ tension related to the mass of KOI-94.01. Recently, \citet{JH22} added TESS photometry to the analysis. Their dynamical fit using four planets (as observed) resulted in three significant masses: 
$m_{94.02}=9.4^{+2.7}_{-2.2}m_\oplus$,
$m_{94.01}=70^{+13}_{-12}m_\oplus$, $m_{94.03}=19\pm3m_\oplus$ 
\footnote{For KOI-94.02 the paper includes a likely typo: the mass originally given was $m_{94.02}=94m_\oplus$, but the bulk density quoted suggests a decimal point was missing.}. We note that their determination of the mass of KOI-94.01 is in between these of \citet{Weiss13} and \citet{masuda13}, consistent with \citet{masuda13} for the other planets. 

Our analysis of the system was able to detect significant planetary masses for these same three planets:
KOI-94.02 / Kepler-89 c (R=$3.777^{+0.034}_{-0.037}R_\oplus$) has a mass of $m_{94.02}=11.6^{+1.6}_{-1.3}m_\oplus$, and
KOI-94.01 / Kepler-89 d (R=$10.242^{+0.091}_{-0.099}R_\oplus$) has a mass of $m_{94.01}=72.9^{+4.5}_{-4.5}m_\oplus$, and
KOI-94.03 / Kepler-89 e (R=$6.110^{+0.055}_{-0.059}R_\oplus$) has a mass of $m_{94.03}=10.49^{+0.98}_{-0.95}m_\oplus$. These values are close to the ones provided by \citet{JH22} and with lower fractional errors, despite the latter having more than three times the temporal baseline available. We note that they deemed their four-planet fit as unsatisfactory (significant residuals) and explored a few scenarios with an additional non-transiting planet. These models naturally improved upon the previous fit's quality, but none of these models was deemed satisfactory, also by \cite{JH22}.

\subsubsection{KOI-520 / Kepler-176}
No literature masses significant to better than $3\sigma$ exist for any planet in the system. We were able to find significant masses for the middle pair of the four planets in the system: 
KOI-520.01 / Kepler-176 c (R=$2.467 \pm 0.025 R_\oplus$) has a mass of $m_{520.01}=1.33^{+0.40}_{-0.27}m_\oplus$, and
KOI-520.03 / Kepler-176 d (R=$2.50 \pm 0.11 R_\oplus$) has a mass of $m_{520.03}=1.72^{+0.46}_{-0.31}m_\oplus$. Both masses result in relatively low densities for planets of this size range. Additionally, significant eccentricity was detected for the innermost planet at $e_{y,520.02}=0.0253^{+0.0082}_{-0.0083}$ and significant relative eccentricity $\Delta e_x=0.0322^{+0.0074}_{-0.0090}$ between the inner pair components. This could possibly account for the somewhat lower mass due to the known mass-eccentricity degeneracy.

\subsubsection{KOI-720 / Kepler-221}

We were able to constrain the masses of the middle pair of this four-planet system:
KOI-720.01 / Kepler-221 c (R=$2.981^{+0.023}_{-0.027}R\oplus$) has a mass of $m_{720.01}=9.3^{+2.4}_{-1.7}m_\oplus$, and
KOI-720.02 / Kepler-221 d (R=$2.813^{+0.022}_{-0.026R}\oplus$) has a mass of $m_{720.02}=3.61^{+0.91}_{-0.81}m_\oplus$. No other significant mass constraints for this system were found in the literature.

\subsubsection{KOI-775 / Kepler-52}

We were able to constrain the masses of the inner pair of this three-planet system.
KOI-775.02 / Kepler-52 b (R=$2.18\pm0.11R\oplus$) has a mass of $m_{775.02}=6.27^{+1.29}_{-0.96}m_\oplus$, and
KOI-775.01 / Kepler-52 c (R=$2.02\pm0.11R\oplus$) has a mass of $m_{775.02}=29.9^{+6.2}_{-4.2}m_\oplus$. It is noteworthy that the innermost planet appears to have significant eccentricity at $e_{x,775.02}=-0.0316^{+0.0058}_{-0.0081}$ and $e_{y,775.02}=-0.083^{+0.012}_{-0.020}$.
The middle planet technically had a significant mass detection by \citet{HL14} but with a derived implausible bulk density of $>44$~gr\,cm$^{-3}$. Other planets in the system did not have a significant literature mass detection.

\subsubsection{KOI-834 / Kepler-238}
The three outermost planets of this five-planet system were found to have significant masses, the first of which has determined mass for the first time. We found that:
KOI-834.02 / Kepler-238 d (R=$3.51\pm0.13R_\oplus$) has a mass of $m_{834.02}=23.1^{+7.9}_{6.1}m_\oplus$, and
KOI-834.01 / Kepler-238 e (R=$8.89\pm0.34R_\oplus$) has a mass of $m_{834.01}=69\pm11m_\oplus$, and
KOI-834.05 / Kepler-238 f (R=$3.41\pm0.13R_\oplus$) has a mass of $m_{834.05}=12.2^{+1.9}_{-1.8}m_\oplus$. \citet{HL17} determined a mass of $M=16.6^{+13.7}_{-3.7}m_\oplus$ for KOI-834.01, which is both suspiciously low for the planet's radius (giving $\rho=0.13$~g~cm$^{-3}$) and unusually asymmetric error ranges. The results provided here for the mass of KOI-834.05  are consistent with- and smaller error ranges than- the literature estimates \citet{Xie14, HL17}.

\subsubsection{KOI-869 / Kepler-245}

Significant mass estimates were only found for the third planet in this four-planet system: KOI-869.03 / Kepler-245 c. Literature mass estimate for this R=$2.502^{+0.051}_{-0.042}R_\oplus$ planet is unreasonably high at 226 $m_\oplus$ \citep{HL14} and therefore ignored. Our analysis suggests a mass of $m_{869.03}=26.7^{+7.0}_{-5.1}m_\oplus$ - which is still unusually (but not unphysically) high.



\subsubsection{KOI-880 / Kepler-82}

This system exhibits four transiting planets, and we were able to constrain the mass of the third member: 
KOI-880.01 / Kepler-82 b (R=$3.821^{+0.071}_{-0.053}R_\oplus$) has a mass of $m_{880.01}=10.0^{+3.3}_{-2.3}m_\oplus$. Early literature sources determined uncharacteristically high masses for this planet ($51-87 m_\oplus$) and statistically significant $\sim19m_\oplus$ mass for the outermost planet \citep{Xie13, HL14}, but later \cite{Freudenthal19} added ground-based observations and found that there is a fifth, outermost and non-transiting planet in the system - Kepler-82f. Taking this planet into account changes the interpretation of the system's TTVs, and indeed \cite{Freudenthal19} found that the mass of the planet in question, KOI-880.01, is $m_{880.01}=12.15^{+0.96}_{-0.87}m_\oplus$ - consistent with our result.

\subsubsection{KOI-952 / Kepler-32}

We were able to constrain the masses of the third and fourth planets in this five-planet system in one of two solutions:
KOI-952.01 / Kepler-32 b (R=$2.233^{+0.074}_{-0.070}R_\oplus$) has a mass of $m_{952.01}=1.72^{+0.58}_{-0.47}m_\oplus$, and 
KOI-952.02 / Kepler-32 c (R=$2.114^{+0.072}_{-0.069}R_\oplus$) has a mass of $m_{952.02}=1.05^{+0.37}_{-0.30}m_\oplus$. Both of these masses are from a single solution, whereas the second solution did not include any significant mass detection. The derived bulk densities of $0.6-0.8\,gr\,cm^{-3}$ are low for the planets' radii, despite that no significant eccentricity was detected anywhere in the system.
No mass was previously significantly constrained in this system (initial detection by \citet{HL14} at $>3\sigma$ significance was later revised and became insignificant \citep{HL17}).



\subsubsection{KOI-1258 / Kepler-281}
We were able to constrain the masses of two of the planets in two distinct solutions:
KOI-1258.02 / Kepler-281 b (R=$3.25^{+0.50}_{-0.48}R_\oplus$) has a mass of $m_{1258.02}=324^{+82}_{-65}m_\oplus$ in solution 1, and no significant mass in solution 2.
KOI-1258.03 / Kepler-281 d (R=$6.3^{+1.3}_{-1.2}R_\oplus$) has a mass of $m_{1258.03}=654^{+100}_{-104}m_\oplus$ in solution 1, and $m_{1258.03}=557^{+93}_{-117}m_\oplus$ in solution 2.

The mass of the inner planet KOI-1258.02 is very large for its radius -- only the large uncertainty on the bulk density allowed this solution to survive -- but we note that the absolute properties of the host star are probably unreliable: e.g, the NASA NExSci database quotes stellar radii between $0.72R_\odot$ \citep{Latham05} to $1.374^{+0.102}_{-0.060}R_\odot$ \citep{2018A&A...616A...1G} for the host star - the latter is from Gaia DR2, whereas the latest Gaia DR3 radius estimate is now $0.880^{+0.027}_{-0.049}R_\odot$ \citep{2023A&A...674A..28F}. We, therefore, posit that there is some normalized geometric and dynamical information in the system in the form of normalized quantities like $r_p/R_s$ and $\mu$, but the absolute sizes and masses are highly uncertain at this stage, possibly due to complicating elements like background/foreground objects or stellar companions. No significant literature masses were found for any of the three planets in this system.

\subsubsection{KOI-1307 / Kepler-287}

No significant literature masses were found for the planets in this system (though \cite{Judkovsky22b} came close). We were able to constrain the masses of both planets to be:
KOI-1307.02 / Kepler-287 b (R=$2.556^{+0.064}_{-0.061}R_\oplus$) has a mass of $m_{1307.02}=83^{+80}_{-26}m_\oplus$,  and 
KOI-1307.01 / Kepler-287 c (R=$2.871^{+0.065}_{-0.060}R_\oplus$) has a mass of $m_{1307.01}=66^{+18}_{-15}m_\oplus$.
The derived bulk densities are too high: only the large uncertainties "save" the solution from being disqualified.

\subsubsection{KOI-1353 / Kepler-289}

This system's history is a bit unusual: currently, there are two confirmed planets with periods of $\sim34$~d and $\sim125$~d, as well as a long-period candidate with a period of $\sim330$~d. However, previously, the same signals were interpreted by a citizen-scientists project as also resulting from a $\sim66$~d candidate \cite{Schmitt2014} that was even given a Kepler-number designation - but this signal is no longer believed to exist in the data. We were able to determine the masses of all three planets in the system, both the innermost and outermost planets having their mass constrained for the first time:
KOI-1353.02 / Kepler-289 b (R=$2.316^{+0.035}_{-0.021}R_\oplus$) has a mass of $m_{1353.02}=24.7^{+7.4}_{-7.9}m_\oplus$,  
KOI-1353.01 / Kepler-289 c (R=$11.47^{+0.17}_{-0.10}R_\oplus$) has a mass of $m_{1353.01}=268^{+30}_{-29}m_\oplus$,  
KOI-1353.03 (R=$3.204^{+0.049}_{-0.029}R_\oplus$) has a mass of $m_{1353.03}=17.4^{+1.3}_{-1.2}m_\oplus$. 

\cite{Santerne2016} analyzed this system using RV and and found that $m_{1353.01} = 1.55\pm{0.35}m_{\jupiter} = 490\pm{110}m_\oplus$,  while using a host mass of $m_{KOI-1353}=1.35^{+0.11}_{-0.07}M\odot$. Note that we used a host mass of $m_{KOI-1353}=1.061^{+0.011}_{-0.035}M\odot$. Correcting the derived RV mass to the same host mass as the one we used in the TTV analysis results in RV mass of:  $m_{1353.01} = 1.22\pm{0.28}m_{\jupiter} = 387\pm{87}m_\oplus$ - which is not far ($\sim1.3\sigma$) from our TTV-derived value, and significantly more massive than the value of $m_{1353.01}=115\pm12$ derived by \citet{JH21}.

\subsubsection{KOI-1529 / Kepler-59}

We were able to determine significant masses for both known planets in this system:
KOI-1529.02 / Kepler-59 b (R=$1.373^{+0.075}_{-0.074}R_\oplus$) has a mass of $m_{1529.02}=4.1^{+2.8}_{-1.0}m_\oplus$, and 
KOI-1529.01 / Kepler-59 c (R=$2.348^{+0.079}_{-0.077}R_\oplus$) has a mass of $m_{1529.01}=3.81^{+2.64}_{-0.98}m_\oplus$. The derived density of KOI-1529.02 ($\rho_{1529.02}=8.7^{+6.0}_{-2.6}~g~cm^{-3}$) is such that the planet appears to be virtually airless, whereas the neighboring planet has bulk density of $\rho_{1529.01}=1.62^{+1.14}_{-0.45}~g~cm^{-3}$ - necessitating a significant atmosphere.

\subsubsection{KOI-1831 / Kepler-324}
No significant literature masses were previously found. We were able to determine the mass of the outer two of the four planets in the system:
KOI-1831.03 / Kepler-324 d (R=$1.5314^{+0.0129}_{-0.0092}R_\oplus$) has a mass of $m_{1831.03}=1.26^{+0.35}_{-0.24}m_\oplus$, and
KOI-1831.01 / Kepler-324 c (R=$2.706^{+0.023}_{-0.016}R_\oplus$) has a mass of $m_{1831.01}=6.2^{+1.7}_{-1.2}m_\oplus$.

\subsubsection{KOI-1977 / Kepler-345}

We found two distinct solutions for this system; both include a significant mass detection for the innermost of this pair of planets:
KOI-1977.02 / Kepler-345 b (R=$0.819\pm0.0630R_\oplus$) has a mass of $m_{1977.02}=1.23^{+0.30}_{-0.26}m_\oplus$ in solution 1, and $m_{1977.02}=0.48^{+0.22}_{-0.14}m_\oplus$ in solution 2. Note that this is a particularly small planet to have its mass significantly measured, with a derived bulk density similar to that of iron in solution 1 and slightly denser than that of Earth in solution 2.


\subsection{Individual systems - literature revision}
In this section, we discuss systems in which none of the significant masses we detected is first in the literature, although some are significantly and interestingly different from previously determined values.

\subsubsection{KOI-115 / Kepler-105}
Only the outermost object in this three-candidate system already got significant literature mass, and we also were only able to constrain the mass of only this planet:
KOI-115.02 / Kepler-105 c (R=$1.655^{+0.025}_{-0.028}R_\oplus$) has a mass of $m_{115.02}=6.71^{+0.71}_{-0.84}m_\oplus$, whereas \citet{JH21} found it to be $M=4.77^{+0.92}_{-0.89}m_\oplus$ - i.e. consistent with our result.

\subsubsection{KOI-137 / Kepler-18}
\cite{Cochran11} modeled the system using early Kepler data (spanning about 500~d), as well as RV and Spitzer photometry. From their adopted model, using all available data, they derived that
KOI-137.01 has a mass of $m_{137.01}=17.3\pm1.9m_\oplus$, and KOI-137.02 has a mass of $m_{137.02}=16.4\pm1.4m_\oplus$. Early analyses based on TTVs \citep{HL14, HL17} arrived at similar masses. More recently, \citet{JH22} added TESS photometry to the full Kepler dataset and found significantly lower masses: $m_{137.01}=6.29^{+3.29}_{-1.59}m_\oplus$ and $m_{137.02}=8.73^{+2.95}_{-1.66}m_\oplus$. Our analysis is consistent with that later analysis, although it converged to two distinct solutions:
KOI-137.01 / Kepler-18 c (R=$4.427^{+0.040}_{-0.022}R_\oplus$) has a mass of $m_{137.01}=6.3^{+2.0}_{-1.3}m_\oplus$ in solution11 and $m_{137.01}=2.76^{+1.09}_{-0.73}m_\oplus$ in solution 2, and
KOI-137.02 / Kepler-18 d (R=$5.623^{+0.051}_{-0.028}R_\oplus$)) has a mass of $m_{137.02}=8.7^{+1.7}_{-1.3}m_\oplus$ in solution 1, and $m_{137.02}=4.7^{+1.4}_{-1.1}m_\oplus$ in solution 2. 

\subsubsection{KOI-152 / Kepler-79}

We constrained the masses of all four planets in the system in two distinct solutions:
KOI-152.03 / Kepler-79 b (R=$3.342^{+0.078}_{-0.043}R_\oplus$)) has a mass of $m_{152.03}=10.3^{+3.2}_{-2.6}m_\oplus$ in solution 1, and $m_{152.03}=15.7^{+3.4}_{-2.5}m_\oplus$ in solution 2.
KOI-152.02 / Kepler-79 c (R=$3.556^{+0.082}_{-0.045}R_\oplus$) has a mass of $m_{152.02}=8.2^{+1.8}_{-1.7}m_\oplus$ in solution 1, and $m_{152.02}=11.4^{+1.4}_{-1.3}m_\oplus$ in solution 2.
KOI-152.01 / Kepler-79 d (R=$6.919^{+0.157}_{-0.082}R_\oplus$) has a mass of $m_{152.01}=10.2^{+1.7}_{-1.6}m_\oplus$ in solution 1, and $m_{152.01}=12.4^{+1.4}_{-1.3}m_\oplus$ in solution 2.
KOI-152.04 / Kepler-79 e (R=$3.42^{+0.15}_{-0.14}R_\oplus$) has a mass of $m_{152.04}=5.97^{+0.74}_{-0.76}m_\oplus$ in solution 1, and $m_{152.04}=6.54^{+0.64}_{-0.61}m_\oplus$ in solution 2.
The system was analyzed multiple times in the past \citep{Xie13, HL14, HL17, JH22}, as well as by us \citep{Yoffe21}. Only the latter analysis was able to detect all masses significantly, and it is fully consistent with the current analysis. Where determinations are available, our results are generally consistent with (but with smaller errors than) the more recent works \citep{HL17, JH22}.

\subsubsection{KOI-156 / Kepler-114}

This three-planet system was studied in the past by several authors, who mostly claimed mass detection of the middle planet, while \citet{Judkovsky22b} claimed the detection of all three masses. We arrived at two distinct solutions, unsurprisingly, one with lower mass and higher eccentricities (solution 1 below) and the other with higher masses and lower eccentricity (solution 2). Our analysis is thus still significantly influenced by the mass-eccentricity degeneracy, and the solutions we arrived at are:
KOI-156.02 / Kepler-114 b (R=$1.332^{+0.062}_{-0.060}R_\oplus$) has a mass of $m_{156.02}=1.09^{+0.49}_{-0.31}m_\oplus$ in  solution 1 and $m_{156.02}=2.46^{+0.86}_{-0.61}m_\oplus$ in solution 2 - a factor of about two.
KOI-156.01 / Kepler-114 c (R=$1.716^{+0.078}_{-0.076}R_\oplus$) has no significant mass in solution 1, and a mass of $m_{156.01}=1.08^{+0.19}_{-0.17}m_\oplus$ in solution 2. Solution 1 is not statistically significant, but its best-fit mass is also a factor of about two of the mass of solution 2.
KOI-156.03 / Kepler-114 d (R=$2.76 \pm 0.18 R_\oplus$) has no significant mass in solution 1, and a mass of $m_{156.03}=1.62^{+0.34}_{-0.33}m_\oplus$ in solution 2, and it too has a factor of about two between the two best-fit masses. The $\Delta e_{x,2}$ component (i.e. relative eccentricity between the inner pair) was $\Delta e^{sol1}_{x,2} = -0.056^{+0.016}_{-0.018}$ in solution 1 and about half as much at $\Delta e^{sol2}_{x,2}=-0.0259^{+0.0059}_{-0.0054}$ in solution 2 - indicating the mass-eccentricity degeneracy. No other significant degeneracies were detected.

For comparison, two past studies found significant masses in the system. The innermost planet was found by \citep{Judkovsky22b} to have a mass of $m_{KOI-156.02}= 3.53^{+0.17}_{-0.15}$, consistent with solution 2. The middle planet, KOI-156.01 was found by \cite{JH21} to have a mass consistent with our result. Notably, but both outer planets are less than half as heavy as determined by \cite{JH21} than by \cite{Judkovsky22b}.

\subsubsection{KOI-157 / Kepler-11}

While the initial analysis of this system suggested different values \citep{Lissauer13} when comparing the results of this study and the results of the most recent analysis of this landmark system \citep[][NSI values]{Bedell17}, there is an overall good agreement across all masses of the inner five planets:
KOI-157.06 / Kepler-11 b (R=$1.890\pm0.023_\oplus$) has a mass of $m_{157.06}=1.07^{+0.36}_{-0.34}m_\oplus$, and
KOI-157.01 / Kepler-11 c (R=$2.955\pm0.034_\oplus$) has a mass of $m_{157.01}=2.08^{+0.74}_{-0.71}m_\oplus$, and
KOI-157.02 / Kepler-11 d (R=$3.244\pm0.038_\oplus$) has a mass of $m_{157.02}=7.47^{+0.47}_{-0.48}m_\oplus$, and
KOI-157.03 / Kepler-11 e (R=$4.287\pm0.050_\oplus$) has a mass of $m_{157.03}=8.79^{+0.56}_{-0.57}m_\oplus$, and
KOI-157.04 / Kepler-11 f (R=$2.614\pm0.035_\oplus$) has a mass of $m_{157.04}=2.12^{+0.31}_{-0.30}m_\oplus$. The mass of KOI-157.01 is very slightly below the $3\sigma$ threshold, and the mass of the outermost planet remains undetermined.

\subsubsection{KOI-209 / Kepler-117}

Our analysis provided constraints on the masses for both planets:
KOI-209.02 / Kepler-117 b (R=$7.97^{+0.25}_{-0.18}R_\oplus$) has a mass of $m_{209.02}=31.4^{+5.4}_{5.7}m_\oplus$, and
KOI-209.01 / Kepler-117 c (R=$12.14^{+0.37}_{-0.28}R_\oplus$) has a mass of $m_{209.01}=808^{+32}_{-34}m_\oplus$. However, TTVs of KOI-209.02 (but not KOI-209.01) also include a second but significant 3:1 component \citep{Ofir18}, which is not included in the current, first-order only model. Indeed, an RV+TTV analysis found a mass for KOI-209.02, which is consistent with our analysis, but $<4\sigma$ lower mass for KOI-209.01 \citet{Bruno15}. The reason that the high-order criterion did not disqualify this system is that the TTVs of KOI-209.02 are unusually-high SNR (as evident by the $>23\sigma$ mass detection of KOI-209.01). While the second-order 3:1 near-MMR signal is not dominant, it is still visible in the data, leading to the tension with \citet{Bruno15}. We reiterate that the high-order criterion checks only for the dominant effect.

\subsubsection{KOI-248 / Kepler-49}
After it was analyzed by \citet{Xie13} and \citet{HL14}, this system's latest analysis was performed by \citet{JH16, JH21} and \cite{HL17}, and the latter constrained the masses of KOI-248.01 and KOI-248.02 -- the two middle planets of this four-planet system -- to better than $3\sigma$.  We were able to determine the masses of the same planets and find the following: 
KOI-248.01 / Kepler-49 b (R=$2.612 \pm 0.079 R_\oplus$) has a mass of $m_{248.01}=5.48^{+0.71}_{-0.80}m_\oplus$, and
KOI-248.02 / Kepler-49 c (R=$2.47 \pm 0.19 R_\oplus$) has a mass of $m_{248.02}=4.95^{+0.75}_{-0.81}m_\oplus$. Both of these values are consistent with- but somewhat lower than- the \citet{HL17} result, as well as with the masses derived by \citet{JH16, JH21} (though these were of lower significance). We note that planets with radii $>2R_\oplus$ rarely have densities similar to Earth's as in the \citet{HL17} solution, thus a lower mass seems more likely.

\subsubsection{KOI-244 / Kepler-25}
An early study of this system found masses for both planets in the system \citet{HL14}, However, the same authors later revised their estimates to lower masses, leaving only the outer planet $m_{244.01}=10.0^{+3.5}_{-2.5}$ with a significant mass detection \citep{HL17}. Later studies did not find significant masses at all \citep[e.g.][]{JH21}). We were able to constrain the mass of the inner planet:
KOI-244.02 / Kepler-25 b (R=$2.757^{+0.019}_{-0.021}R\oplus$) has a mass of $m_{244.02}=1.63^{+1.40}_{-0.54}m_\oplus$, corresponding to a density of  $0.43^{+0.37}_{-0.14} g cm^{-3}$ - which would make it a member of the little-understood group of low-mass low-density planets.

\subsubsection{KOI-274 / Kepler-122}
We were able to find significant mass for both planets in this two-planet system in two separate solutions:
KOI-274.01 / Kepler-128 b (R=$1.539^{+0.039}_{-0.036}R_\oplus$) has a mass of $m_{274.01}=9.1^{+2.6}_{-1.5}m_\oplus$ in solution 1, and $m_{274.01}=2.47^{+1.13}_{-0.51}m_\oplus$ in solution 2. 
KOI-274.02 / Kepler-128 c (R=$1.379^{+0.043}_{-0.041}R_\oplus$) has a mass of $m_{274.02}=8.4^{+2.3}_{-1.4}m_\oplus$ in solution 1, and $m_{274.02}=2.23^{+1.02}_{-0.46}m_\oplus$ in solution 2. Both planets technically had significant masses detected before \citep{HL14, Xie14} - but those were of unreasonable densities (in the range of 46-152~g~cm$^{-3}$), and more recent studies \citep{HL16, HL17}  indeed found smaller masses - but those were not $3\sigma$ significant. Solution 1 above also results in high bulk densities but not large enough to be disqualified, whereas solution 2 results in reasonable densities. As expected, the $\sim3$ fold decrease in mass between the two solutions was accompanied by a $\sim3$ fold increase in $\Delta e_x$.

\subsubsection{KOI-277 / Kepler-36}
We were able to find significant mass for both planets in this two-planet system:
KOI-277.02 / Kepler-36 b (R=$1.584^{+0.025}_{-0.024}R_\oplus$) has a mass of $m_{277.02}=6.41^{+0.21}_{-0.23}m_\oplus$, and
KOI-277.01 / Kepler-36 c (R=$3.696^{+0.047}_{-0.044}R_\oplus$) has a mass of $m_{277.01}=11.59^{+0.45}_{-0.40}m_\oplus$. Both masses are significantly higher than in previous analyses \citep[e.g.][]{Carter2012, Vissapragada20, JH21}, but the mass ratio of the two planets is virtually identical to the one found by these studies. Both the absolute and the relative eccentricity components between the planets are low (consistent with zero to less than $1\sigma$), making the system particularly vulnerable to the effects of the mass-eccentricity degeneracy.
The observed typical differences in masses relative to the literature of $\sim30\%$ would correspond to a similar compensating relative deviation in eccentricity. Our analysis shows these are statistically indistinguishable from our results, which are consistent with zero at the $1\sigma$ level.

\subsubsection{KOI-370 / Kepler-145}

We found significant masses for both planets in the system: 
KOI-370.02 / Kepler-145 c (R=$2.44\pm0.20R_\oplus$) has a mass of $m_{370.02}=34^{+25}_{-11}m_\oplus$, and
KOI-370.01 / Kepler-145 b (R=$3.77\pm0.16R_\oplus$) has a mass of $m_{370.01}=25.9^{+9.4}_{-6.0}m_\oplus$. However, these masses are too high to be likely. This system was also analyzed by \citet{Xie14}, who also found similarly high masses. The recent analysis by \citet{Judkovsky22b} produced a lower mass of $\sim20m_\oplus$  for KOI-370.02 - which still implies a density of $7.57\pm0.62 g\,cm^{-3}$ - higher than expected for a planet in this size range. The combination of consistently high masses (across multiple studies) demands further study.

\subsubsection{KOI-457 / Kepler-161}
We were able to constrain the masses of both planets in the system in two distinct solutions:
KOI-457.01 / Kepler-161 b (R=$2.237^{0.014}_{-0.023}R_\oplus$) has a mass of $m_{457.01}=2.63^{+0.83}_{-0.61}m_\oplus$ in solution 1, and an insignificant mass detection in solution 2.
KOI-457.02 / Kepler-161 c (R=$2.476^{0.014}_{-0.025}R_\oplus$) has a mass of $m_{457.02}=5.65^{+0.91}_{-1.12}m_\oplus$ in solution 1, and $m_{457.02}=3.61^{+0.78}_{-1.19}m_\oplus$ in solution 2. As expected, solution 2 includes a higher (and statistically significant) eccentricity component: $\Delta e_x=0.065^{+0.021}_{-0.019}$, whereas solution 1 does not.
\citet{Judkovsky22a} also detected significant masses in this system: $m_{457.01}=4.43^{+0.37}_{-0.40}$  and $m_{457.02}=4.49^{+0.96}_{-0.85}$, i.e. consistent with solution 1 above.

\subsubsection{KOI-500 / Kepler-80}

A five-planet dynamical analysis was performed by \citet{MacDonald16}, but later, a sixth set of transit signals was detected \citep{Shallue18} with a period of $\sim14.6$d - the outermost known planet in the system. This last planet is very small and has not been found to have a significant mass in our analysis. We therefore proceed with a comparison with \citet{MacDonald16}, despite the lack of the sixth planet. We found two sets of solutions, with significant masses for the fourth and fifth planets in both: 
KOI-500.01 / Kepler-80 b (R=$2.576^{+0.024}_{-0.030}R_\oplus$) has a mass of $m_{500.01}=0.57^{+0.30}_{-0.17}m_\oplus$ in solution 1, and , $m_{500.01}=1.74^{+0.50}_{-0.37}m_\oplus$ in solution 2.
KOI-500.02 / Kepler-80 c (=$2.736^{+0.030}_{-0.035}R_\oplus$) has a mass of $m_{500.02}=1.11^{+0.58}_{-0.34}m_\oplus$ in solution 1, and $m_{500.02}=3.45^{+0.91}_{-0.80}m_\oplus$ in solution 2. While solution 2 appears more plausible, we note that $\Delta e_y$ between these two planets, and only this eccentricity component, is significant in both solutions -- and it decreased from $-0.060^{+0.020}_{-0.023}$ in solution 1 to $-0.0128^{+0.0040}_{-0.0047}$ in solution 2. Therefore the system appears to be still heavily influenced by the mass-eccentricity degeneracy.

Moreover, a detailed analysis by \citet{MacDonald16} showed that the system includes a rare dynamical configuration with four interconnected three-body resonances and, consequently, significantly different derived masses. The system was not flagged as being in-MMR since the tools presented here to avoid simple two-body MMRs are not suitable for this rare multi-body resonance.

\subsubsection{KOI-547 / Kepler-595}

Significant masses in this system were previously detected only by \citet{Yoffe21}, using the same dynamical core as the current analysis. We are now able to refine our results and find two distinct solutions:
KOI-547.03 / Kepler-595 c (R=$1.057^{+0.093}_{-0.092}R_\oplus$) has a mass of $m_{547.03}=3.06^{+1.09}_{-0.53}m_\oplus$ in solution 1, and a mass of $m_{547.03}=1.48^{+0.45}_{-0.30}m_\oplus$ in solution 2, and 
KOI-547.01 / Kepler-595 b (R=$3.76^{+0.18}_{-0.17}R_\oplus$) has a mass of $m_{547.01}=23.6^{+6.0}_{-4.0}m_\oplus$ in solution 1, and a mass of $m_{547.01}=11.9^{+3.3}_{-2.9}m_\oplus$. The first solution is virtually identical to our previous analysis \citep{Yoffe21}, and the second solution exhibits planets at about half the mass and twice the magnitude of both eccentricity components (though with uncertainties comparable to the magnitude).

\subsubsection{KOI-620 / Kepler-51}

This three-planet system is home to some of the lowest-density planets known. Our best-fit solution provided masses of the outer two planets:
KOI-620.03 / Kepler-51 c (R=$5.81\pm0.20R_\oplus$) has a mass of $m_{620.03}=3.73^{+0.19}_{-0.17}m_\oplus$, 
KOI-620.02 / Kepler-51 d (R=$9.08\pm0.14R_\oplus$) has a mass of $m_{620.02}=6.79^{+0.93}_{-0.90}m_\oplus$.
Both of these are consistent with literature values \citep{HL17, LibbyRoberts20, JH22} with significantly smaller error ranges in the case of KOI-620.03. 

\subsubsection{KOI-829 / Kepler-53}

Significant masses were found in the literature for the outer pair of this three-planet system \citep{}{most recently by }{HL17}. Our analysis was able to find masses for these same planets in two distinct solutions, both of which are consistent with the literature and with significantly reduced uncertainties:
KOI-829.01 / Kepler-53 b (R=$3.53^{+0.16}_{-0.15}R_\oplus$) has a mass of $m_{829.01}=17.8^{+4.0}_{-2.6}m_\oplus$ in solution 1 and $m_{829.01}=9.5^{+3.3}_{-2.7}m_\oplus$ in solution 2, while
KOI-829.03 / Kepler-53 c (R=$3.82^{+0.18}_{-0.16}R_\oplus$) has a mass of $m_{829.03}=16.3^{+2.6}_{-2.0}m_\oplus$ in solution 1, and $m_{829.03}=9.3^{+2.8}_{-2.4}m_\oplus$ in solution 2.

\subsubsection{KOI-886 / Kepler-54}

We derived constraints on the masses of the inner pair of this three-planet system:
KOI-886.01 / Kepler-54 b ((R=$1.921 \pm 0.062 R_\oplus$) has a mass of $m_{886.01}=0.98^{+0.34}_{-0.22}m_\oplus$, and
KOI-886.02 / Kepler-54 c (R=$1.444 \pm 0.062 R_\oplus$) has a mass of $m_{886.02}=0.70^{+0.25}_{-0.16}m_\oplus$. Both of these masses are low for the planets' radii. Looking deeper into the system, we find that the system is close to residing in MMR, with a normalized distance from resonance of $\Delta=0.00459$. Systems so close to MMR can become caught in the resonance given high enough masses, and indeed, most of our solutions for this system resulted in high masses and were discarded for the in-resonance criterion. To further evaluate if the system is near or within MMR, we examine the observed TTVs of the two planets. We find they share clear and anti-correlated TTVs with a similar period of about $850\pm5$~d \citet{Holczer16}. This is close to the super-period predicted by the near-MMR interaction \citet{LithXieWu12} of $876$~d, while the in-MMR libration period is predicted to be more than an order of magnitude longer, at about $10,991$~d [eq. 8.47 of \citet{Murray2000}]. Thus, this characteristic timescale favors near-MMR dynamics as opposed to in-MMR behaviour.

Looking in the literature, significant mass detections were claimed by \cite{HL14, HL17}, and between the two analyses, the claimed masses decreased by a factor $>8$. Later, \citet{JH21} estimated the expected TTVs for different systems comparing cases where they are in- or just near- MMR and KOI-886 received a much higher "Resonant TTV Score" than "Nonresonant TTV Score" - by factors of $>6$ and $>18$ for the two planets. \citet{JH21} could not detect any mass in the system to more than $3\sigma$ and indeed suggested that more data is required to properly analyze the system. We conclude that the KOI-886 system appears to lie just at the border of the validity of our analysis: close to- but outside of- the 3:2 MMR, and our solution above is the best one that does not violate the other acceptance criteria. However, this solution may still suffer from reduced precision due to the limitations of \texttt{TTVFaster} too close to resonance (see Figure 4 of \cite{TTVFaster}).

\subsubsection{KOI-898 / Kepler-83}

We found significant masses for all three planets in the system:
KOI-898.02 / Kepler-83 d (R=$2.006 \pm 0.072 R_\oplus$) has a mass of $m_{898.02}=4.1^{+2.8}_{-1.3}m_\oplus$.
KOI-898.01 / Kepler-83 b (R=$2.689 \pm 0.091 R_\oplus$) has a mass of $m_{898.01}=2.94^{+0.88}_{-0.62}m_\oplus$, and
KOI-898.03 / Kepler-83 c (R=$2.71 \pm 0.27 R_\oplus$) has a mass of $m_{898.03}=2.03^{+0.52}_{-0.39}m_\oplus$.
These values are slightly lower than the only other significant literature masses \citep{Judkovsky22b}. Technically, we also found another valid solution for the system, but none of the masses was significant in that solution.

\subsubsection{KOI-935 / Kepler-31}
In this system, the innermost candidate is still not validated, while the outer three already are. Of these three, we provide a mass estimate for the middle planet:
KOI-935.02 / Kepler-31 c (R=$5.05^{+0.19}_{-0.18}R_\oplus$) has a mass of $m_{935.02}=17.1^{+2.5}_{-5.6}m_\oplus$ (consistent with literature values by \citet{HL17} in both value and error).

\subsubsection{KOI-1203 / Kepler-276}

Our best fit for the system includes a significant mass for the middle planet in this three-planet system:
KOI-1203.01 / Kepler-276 c (R=$2.853^{+0.108}_{-0.095}R_\oplus$) has a mass of $m_{1203.01}=3.7^{+2.1}_{-1.2}m_\oplus$. This value is in tension with the higher masses given in the current literature \citet{Xie14, Judkovsky22b} to the level of $3.1\sigma$ and $3.2\sigma$, respectively. These higher masses would result in a rocky bulk density for the planet, even though most planets in that radius range usually have substantial atmospheres.

\subsubsection{KOI-1102 / Kepler-24}
The middle pair of this four-planet system have reported significant masses \citep{HL17}:
KOI-1102.02 / Kepler-24 b (R=$2.404\pm0.063R_\oplus$) was estimated to have a mass of $m_{1102.02}=7.5^{+2.3}_{-2.4}m_\oplus$ and we now estimate it to have a mass of $m_{1102.02}=8.5^{+3.2}_{-2.0}m_\oplus$,  and
KOI-1102.01 / Kepler-24 c (R=$2.544\pm0.067R_\oplus$) was estimated to have a mass of $m_{1102.01}=5.2^{+2.0}_{-1.4}m_\oplus$ and we now estimate it to have a mass of $m_{1102.01}=6.9^{+2.8}_{-1.6}m_\oplus$, showing agreement for both planets.

\subsubsection{KOI-1215 / Kepler-227}

We were able to determine significant masses for both planets in the system:
KOI-1215.01 / Kepler-227 b (R=$2.721^{+0.046}_{-0.042}R_\oplus$) has a mass of $m_{1215.01}=33.6^{+15.1}_{-8.2}m_\oplus$, and
KOI-1215.02 / Kepler-227 c (R=$3.061^{+0.054}_{-0.048}R_\oplus$) has a mass of $m_{1215.02}=40.4^{+9.4}_{-7.0}m_\oplus$. These masses appear to be unusually high (albeit uncertain) for the planets' radii. Interestingly, this system was also analyzed by no fewer than four other studies \citep{Xie14, HL14, HL17, Judkovsky22b} - which all found similarly high (or higher still) masses. The systematically high masses are currently unexplained, and we speculate that additional non-transiting components bias all TTV analyses.

\subsubsection{KOI-1236 / Kepler-279}

This system includes four sets of transit signals. The first three are confirmed planets, whereas the outermost signal, with a period of $P_{1236.04}\approx98.35d$, was considered a candidate. We are able to constrain the masses of the outer three planets, i.e., also to confirm the outermost candidate as a real planet:
KOI-1236.01 / Kepler-279 c (R=$4.62 \pm 0.37 R_\oplus$) has a mass of $m_{1236.01}=18.1^{+3.2}_{-2.7}m_\oplus$, 
KOI-1236.03 / Kepler-279 d (R=$4.29^{+0.40}_{-0.39}R_\oplus$) has a mass of $m_{1236.03}=9.5^{+1.6}_{-1.4}m_\oplus$, 
KOI-1236.04 - hereafter also Kepler-279 e - (R=$3.48 \pm 0.54 R_\oplus$) has a mass of $m_{1236.04}=56.0^{+6.0}_{-4.9}m_\oplus$. We note that the innermost planet has a surprisingly large best-fit mass, but due to its large errors, it is statistically insignificant.

In the past, the middle pair of this system was analyzed, with \cite{Xie14} estimating masses about $\sim3$ times higher and \cite{HL17} estimating masses $\sim2$ times lower than our estimates above.

\subsubsection{KOI-1783 / Kepler-1662}
Of the two known planets, we determined the mass of the inner giant planet:
KOI-1783.01 / Kepler-1662 b (R=$9.01^{+0.19}_{-0.18}R_\oplus$) has a mass of $m_{1783.01}=283^{+87}_{-47}m_\oplus$. This is significantly heavier than $m_{1783.01}=71.0^{+11.2}_{-9.2}$ found by \citet{Vissapragada20} or $m_{1783.01}=99^{+33}_{-25}$ of \cite{JH21}.

\subsubsection{KOI-2025 / Kepler-350}

We were able to determine the mass of all three planets in the system in one of two solutions:
KOI-2025.03 / Kepler-350 b (R=$1.76 \pm 0.10 R_\oplus$) has a mass of $m_{2025.03}=16.9^{+5.6}_{-3.3}m_\oplus$, 
KOI-2025.01 / Kepler-350 c (R=$3.19^{+0.32}_{-0.31}R_\oplus$) has a mass of $m_{2025.01}=3.80^{+1.00}_{-0.77}m_\oplus$, and
KOI-2025.02 / Kepler-350 d (R=$2.72^{+0.15}_{-0.14}R_\oplus$) has a mass of $m_{2025.02}=9.8^{+1.7}_{-1.1}m_\oplus$.  All these masses are from a single solution, whereas the second solution does not include any significant mass detection. 

Older studies of this system \citep{Xie14, HL14} were partial (only the outer pair of this three-planet system) and with either insignificant or marginally significant mass detections. More recently, \citet{Judkovsky22b} analyzed the full system, and their results regarding the outer pair are statistically similar to results presented here, but there is tension at the $4\sigma$ level regarding the innermost planet: they determined its mass to be $M=3.65^{+0.35}_{-0.38}m_\oplus$.

\section{Conclusions}
\label{conclusions}


We presented an application of the \texttt{PyDynamicaLC} code to the entire \textit{Kepler} sample of multiple systems, refined our interpretation of the results, and were, therefore, able to constrain the masses of \SignificantMass planets. \NewSignificantMass of these mass estimates are better than $3\sigma$ for the first time in the literature. The goal of \texttt{PyDynamicaLC} is to determine the masses of the smallest and lightest possible planets, and indeed, we were able to significantly determine the masses of \SmallPlanets planets with radii smaller than $2R_\oplus$, \NewSmallPlanets of them for the first time, and down to less than $1R_\oplus$ for a few objects (all in at least one valid solution). These are significant additions to the set of Kepler's small planets with known masses. All these were achieved by tailoring the analysis to systems that are dynamically cool, thereby simplifying the analysis to only first-order effects with a minimal number of degrees of freedom. Additionally, the choices to model the light curve (and not the times of mid-transit) as well as of the model parameters - were all made to maximize the sensitivity of the analysis, even at the cost of generality. In a few cases (KOI-209, KOI-277), we saw the limitation of a first-order only model when higher-order contributions became discernible in high-SNR systems. We recall that the goal of this work is to constrain the lowest-amplitude TTVs and not the detailed modeling of high-quality TTV signals such as these above cases, so this tradeoff is expected.

In the future, better and much longer baseline data sets will become available ({\it e.g.} from the \textit{TESS} and \textit{PLATO} missions). These will allow the detection of ever smaller planets, and measuring their masses will become ever more challenging. Analyzing these future datasets, among others, with \texttt{PyDynamicaLC} will allow constraining the masses of the smallest planets around the brightest stars without resorting to expensive additional resources like radial velocity measurements.

\section*{Acknowledgements}

We thank the anonymous referee for providing many insightful comments that improved this paper, throughout the long process involved.
This study was supported by the Helen Kimmel Center for Planetary Sciences and the Minerva Center for Life Under Extreme Planetary Conditions \#13599 at the Weizmann Institute of Science. The \textit{Kepler} data is available at MAST: \dataset[10.17909/T9488N]{\doi{10.17909/T9488N}}. {A.O. acknowledges the support of the Stephen and Claire Reich Research fellow chair in Chemistry.} Some of the data presented in this paper were obtained from the Mikulski Archive for Space Telescopes (MAST) at the Space Telescope Science Institute. The specific observations analyzed can be accessed via \\dataset[10.17909/T9488N] {https://doi.org/10.17909/T9488N}.

\def\mnras{Month. Not. Royal Acad. Soc.}
\def\apj{Astrophys. J.}
\def\aap{Astro. Astrophys.}
\def\apjl{Astrophys. J. Let.}
\def\physrep{Physics Reports}
\def\aj{Astronomical Journal}
\def\apjs{Astrophys. J., Supp.}
\def\nat{Nature}
\def\icarus{Icarus}
\def\solphys{Solar Physics}
\def\araa{Ann. Rev. of Astron and Astrophys}

\newpage

\newcommand{\ch}[1]{\hspace*{\fill}#1\hspace*{\fill}}

\begin{table*}
\tiny
\centering
\begin{tabular}{| l |r | r | r | c | c |c |c  |c  |c | } 
\hline
KOI host& Mass & Mass Err& Mass err& Radius& Radius Err& Radius err& LDcoeff1& LDcoeff2& Provenance               \\
     &$M_\odot$&$M_\odot$&$M_\odot$&$R_\odot$&$R_\odot$&$R_\odot$  &         &         &                          \\
\hline
94      & 1.177& 0.032   & 0.024   & 1.352 & 0.0120    & 0.0130    & 0.320   & 0.305   & 1 \\
115     & 0.991& 0.032   & 0.033   & 1.023 & 0.0130    & 0.0150    & 0.393   & 0.268   & 1 \\
137     & 0.983& 0.026   & 0.024   & 0.903 & 0.0080    & 0.0040    & 0.526   & 0.186   & 1 \\
152     & 1.244& 0.027   & 0.042   & 1.283 & 0.0290    & 0.0150    & 0.320   & 0.305   & 1 \\
156     & 0.560& 0.031   & 0.031   & 0.715 & 0.0290    & 0.0290    & 0.464   & 0.259   & 3 \\
157     & 0.986& 0.034   & 0.033   & 1.064 & 0.0120    & 0.0120    & 0.404   & 0.262   & 1 \\
209     & 1.180& 0.037   & 0.052   & 1.555 & 0.0480    & 0.0360    & 0.317   & 0.303   & 1 \\
244     & 1.148& 0.035   & 0.033   & 1.334 & 0.0090    & 0.0100    & 0.319   & 0.302   & 1 \\
248     & 0.541& 0.037   & 0.037   & 0.625 & 0.0180    & 0.0180    & 0.422   & 0.304   & 3 \\
274     & 1.155& 0.073   & 0.041   & 1.653 & 0.0210    & 0.0150    & 0.347   & 0.291   & 1 \\
 \hline
 \end{tabular}
 \caption{Stellar parameters used in this work. The Err and err column stand for the relevant parameter uncertainties in the positive and negative directions, respectively, and the LDcoeffs are the limb darkening coefficients used in the quadratic limb darkening law in the light curve fit. Provenances are: 1: \citet{Fulton2018}, 2: NExSci Table, 3: NExSci Table (mass) and \cite{Berger18} (radius). The full table is given in a machine-readable format, and these are just the first few lines to inform the reader of the form and content of the table.}
 \label{StellarParams}
\end{table*}

\begin{table*}
\tiny
 \centering
 \begin{tabular}{| l |c | c | c | c | c |c |c  |c  |c | c |} 
 \hline
KOI     & period         & period Err    & ${T_{\rm mid}}$ & ${T_{\rm mid}}$ Err& $a/R$ & $a/R$ Err&  $b/R$& $b/R$ Err& $r/R$& $r/R$ Err\\
        & [d]            & [d]           & [KBJD]          & [KBJD]             &       &          &       &          &      &          \\
 \hline
94.04   &    3.7431667&   0.0000056&      131.6197&      0.0014&     7.85&    0.18&     0.01&    0.20&    0.01031&   0.00035\\
94.02   &   10.4236774&   0.0000072&     138.00904&     0.00058&    15.64&    0.22&     0.01&    0.15&    0.02556&   0.00018\\
94.01   &   22.3429648&   0.0000034&     132.74164&     0.00013&    27.37&    0.12&    0.031&   0.038&   0.068705&  0.000047\\
94.03   &    54.319962&    0.000034&     161.23987&     0.00050&       44&      17&     0.50&    0.18&    0.04178&   0.00052\\
115.03  &     3.435916&    0.000047&       132.662&       0.011&      6.7&     3.0&     0.62&    0.26&    0.00461&   0.00038\\
115.01  &    5.4122039&   0.0000031&     133.14309&     0.00047&      9.2&     4.5&     0.79&    0.28&    0.02477&   0.00097\\
115.02  &     7.125944&    0.000014&      139.0049&      0.0016&      8.6&     4.2&     0.90&    0.36&     0.0151&    0.0013\\
137.03  &    3.5046913&   0.0000031&     133.51194&     0.00071&      9.5&     3.9&     0.74&    0.28&     0.0176&    0.0016\\
137.01  &    7.6415676&   0.0000018&     135.40756&     0.00019&   17.852&   0.024&    0.027&   0.097&    0.04263&   0.00016\\
137.02  &   14.8589085&   0.0000038&     128.15487&     0.00022&     32.9&     3.3&     0.26&    0.14&    0.05159&   0.00032\\
152.03  &    13.484577&    0.000023&      136.6181&      0.0013&    21.01&    0.71&     0.01&    0.17&    0.02345&   0.00028\\

 \hline
 \end{tabular}
 \caption{Planetary Mandel-Agol parameters used in this work. The Err columns stand for the relevant parameter uncertainty. The full table is given in a machine-readable format, and above are just the first few lines to inform the reader of the form and content of the table.}
 \label{PlanetGeometricParams}
\end{table*}

\tiny
\begin{table*}
\tiny
 \centering
 \begin{tabular}{| r |r | r | r | r | r | r | r | r |r |r  |r  |r |r |r | r |r |} 
\hline
KOI     & Soln.& $\mu\cdot 10^6$ & $\mu$ Err   & $\mu$ err   & Mass         & Mass Err     & Mass err      & $\Delta(e_x)$ & $\Delta(e_x)$ Err & $\Delta(e_x)$ err & $\Delta(e_y)$ & $\Delta(e_y)$ Err  & $\Delta(e_y)$ err  & $\rho$         & $\rho$ Err    & $\rho$ err    \\
\,        & \,  \#          & \,                & \,            & \,            & $[m_\oplus]$ & $[m_\oplus]$ & $[m_\oplus]$  &  \,             & \,                  & \,                  & \,              & \,                   & \,                   & [g cm$^{-3}$]  & [g cm$^{-3}$] & [g cm$^{-3}$] \\
\hline
94.04   & 1           &              75 &       +1309 &         -74 &      1648.85 &            0 &             0 &        0.0084 &           +0.0067 &           -0.0053 &        0.0150 &            +0.0052 &            -0.0081 &             44 &          +765 &           -43 \\
94.02   & 1           &            29.7 &        +4.2 &        -3.4 &         11.6 &         +1.6 &          -1.3 &        0.0182 &           +0.0051 &           -0.0062 &        0.0024 &            +0.0086 &            -0.0062 &           1.19 &         +0.17 &         -0.14 \\
94.01   & 1           &             186 &         +11 &         -12 &         72.9 &         +4.5 &          -4.5 &       -0.0048 &           +0.0051 &           -0.0051 &        0.0470 &            +0.0042 &            -0.0040 &          0.374 &        +0.025 &        -0.025 \\
94.03   & 1           &            26.7 &        +2.5 &        -2.4 &        10.49 &        +0.98 &         -0.95 &       -0.0151 &           +0.0056 &           -0.0066 &       -0.0093 &            +0.0078 &            -0.0081 &          0.253 &        +0.025 &        -0.024 \\
115.03  & 1           &             6.8 &       +17.3 &        -6.5 &      18.1969 &            0 &             0 &       -0.0007 &           +0.0086 &           -0.0083 &        0.0018 &            +0.0082 &            -0.0086 &             72 &          +184 &           -69 \\
115.01  & 1           &             3.2 &        +6.8 &        -2.9 &      10.6263 &            0 &             0 &       -0.0099 &           +0.0093 &           -0.0063 &        0.0115 &            +0.0083 &            -0.0143 &           0.27 &         +0.57 &         -0.24 \\
115.02  & 1           &            20.3 &        +2.2 &        -2.5 &         6.71 &        +0.71 &         -0.84 &       -0.0005 &           +0.0015 &           -0.0018 &       -0.0133 &            +0.0024 &            -0.0017 &           8.16 &         +0.96 &         -1.08 \\
137.03  & 1           &              97 &        +120 &         -50 &      140.397 &            0 &             0 &       -0.0062 &           +0.0045 &           -0.0067 &       -0.0139 &            +0.0072 &            -0.0056 &             34 &           +42 &           -17 \\
137.01  & 1           &            19.1 &        +6.1 &        -3.9 &          6.3 &         +2.0 &          -1.3 &        0.0106 &           +0.0067 &           -0.0046 &       -0.0187 &            +0.0083 &            -0.0098 &          0.398 &        +0.128 &        -0.083 \\
137.02  & 1           &            26.5 &        +5.1 &        -4.0 &          8.7 &         +1.7 &          -1.3 &       -0.0062 &           +0.0053 &           -0.0058 &       -0.0006 &            +0.0060 &            -0.0058 &          0.269 &        +0.052 &        -0.042 \\
137.03  & 2           &             9.4 &       +13.0 &        -8.7 &      30.4165 &            0 &             0 &        -0.020 &            +0.022 &            -0.019 &        -0.022 &             +0.019 &             -0.026 &            3.3 &          +4.5 &          -3.0 \\
137.01  & 2           &             8.4 &        +3.3 &        -2.2 &         2.76 &        +1.09 &         -0.73 &         0.031 &            +0.019 &            -0.022 &        -0.066 &             +0.025 &             -0.023 &          0.176 &        +0.070 &        -0.047 \\
137.02  & 2           &            14.2 &        +4.4 &        -3.3 &          4.7 &         +1.4 &          -1.1 &        -0.014 &            +0.011 &            -0.013 &        -0.001 &             +0.016 &             -0.013 &          0.145 &        +0.045 &        -0.033 \\
 \hline
 \end{tabular}
 \caption{Dynamical parameters found in this work. The Solution \# column allows differentiating between different solutions of the same system (as described in the text, there are up to two solutions for all systems). In all cases $\mu$ states the usual value and $1\sigma$ uncertainties, However, the mass column is slightly different: for significant detections $\mathrm{Mass}>3\,\mathrm{Mass_{err}}$ the mass and $1\sigma$ uncertainties are given, while for masses that are not statistically significant the value given is the $3\sigma$ upper limit, and the uncertainties are set to zero. $\Delta e_x$ and $\Delta e_y$ for each planet are the differences in eccentricity vector components between each planet and the one interior to it in the system. The $\Delta e_x$ and $\Delta e_y$ values for the innermost planet in each system are understood to be the magnitude of the same eccentricity components for that innermost planet. The columns designated with Err (err) are the uncertainty of the corresponding parameter in the positive (negative) direction, respectively. The full table is given in a machine-readable format, and the above are just the first few lines to inform the reader of the form and content of the table.}
 \label{PlanetParams}
\end{table*}

\bibliography{myref}

\begin{thebibliography}{}
\makeatletter
\relax
\def\mn@urlcharsother{\let\do\@makeother \do\$\do\&\do\#\do\^\do\_\do\%\do\~}
\def\mn@doi{\begingroup\mn@urlcharsother \@ifnextchar [ {\mn@doi@}
  {\mn@doi@[]}}
\def\mn@doi@[#1]#2{\def\@tempa{#1}\ifx\@tempa\@empty \href
  {http://dx.doi.org/#2} {doi:#2}\else \href {http://dx.doi.org/#2} {#1}\fi
  \endgroup}
\def\mn@eprint#1#2{\mn@eprint@#1:#2::\@nil}
\def\mn@eprint@arXiv#1{\href {http://arxiv.org/abs/#1} {{\tt arXiv:#1}}}
\def\mn@eprint@dblp#1{\href {http://dblp.uni-trier.de/rec/bibtex/#1.xml}
  {dblp:#1}}
\def\mn@eprint@#1:#2:#3:#4\@nil{\def\@tempa {#1}\def\@tempb {#2}\def\@tempc
  {#3}\ifx \@tempc \@empty \let \@tempc \@tempb \let \@tempb \@tempa \fi \ifx
  \@tempb \@empty \def\@tempb {arXiv}\fi \@ifundefined
  {mn@eprint@\@tempb}{\@tempb:\@tempc}{\expandafter \expandafter \csname
  mn@eprint@\@tempb\endcsname \expandafter{\@tempc}}}

\bibitem[\protect\citeauthoryear{Agol \& Deck}{Agol \& Deck}{2016}]{TTVFaster}
Agol E.,  Deck K.,  2016, \mn@doi [\apj] {10.3847/0004-637x/818/2/177}, 818,
  177

\bibitem[\protect\citeauthoryear{Agol, Steffen, Sari  \& Clarkson}{Agol
  et~al.}{2005}]{Agol05}
Agol E.,  Steffen J.,  Sari R.,   Clarkson W.,  2005, \mn@doi [\mnras]
  {10.1111/j.1365-2966.2005.08922.x}, 359, 567

\bibitem[\protect\citeauthoryear{{Bedell} et~al.,}{{Bedell}
  et~al.}{2017}]{Bedell17}
{Bedell} M.,  et~al., 2017, \mn@doi [\apj] {10.3847/1538-4357/aa6a1d}, \href
  {https://ui.adsabs.harvard.edu/abs/2017ApJ...839...94B} {839, 94}

\bibitem[\protect\citeauthoryear{{Berger}, {Huber}, {Gaidos}  \& {van
  Saders}}{{Berger} et~al.}{2018}]{Berger18}
{Berger} T.~A.,  {Huber} D.,  {Gaidos} E.,   {van Saders} J.~L.,  2018, \mn@doi
  [\apj] {10.3847/1538-4357/aada83}, \href
  {https://ui.adsabs.harvard.edu/abs/2018ApJ...866...99B} {866, 99}

\bibitem[\protect\citeauthoryear{Braak}{Braak}{2006}]{Braak06}
Braak C. J. F.~T.,  2006, \mn@doi [Statistics and Computing]
  {10.1007/s11222-006-8769-1}, 16, 239

\bibitem[\protect\citeauthoryear{{Bruno} et~al.,}{{Bruno}
  et~al.}{2015}]{Bruno15}
{Bruno} G.,  et~al., 2015, \mn@doi [\aap] {10.1051/0004-6361/201424591}, \href
  {https://ui.adsabs.harvard.edu/abs/2015A&A...573A.124B} {573, A124}

\bibitem[\protect\citeauthoryear{Carter et~al.,}{Carter
  et~al.}{2012}]{Carter2012}
Carter J.~A.,  et~al., 2012, \mn@doi [Science] {10.1126/science.1223269}, 337,
  556

\bibitem[\protect\citeauthoryear{{Cochran} et~al.,}{{Cochran}
  et~al.}{2011}]{Cochran11}
{Cochran} W.~D.,  et~al., 2011, \mn@doi [\apjs] {10.1088/0067-0049/197/1/7},
  \href {https://ui.adsabs.harvard.edu/abs/2011ApJS..197....7C} {197, 7}

\bibitem[\protect\citeauthoryear{Feroz, Hobson  \& Bridges}{Feroz
  et~al.}{2008}]{Multinest}
Feroz F.,  Hobson M.~P.,   Bridges M.,  2008, \mn@doi [Mon. Not. Roy. Astron.
  Soc. 398: 1601-1614,2009] {10.1111/j.1365-2966.2009.14548.x}

\bibitem[\protect\citeauthoryear{{Fouesneau} et~al.,}{{Fouesneau}
  et~al.}{2023}]{2023A&A...674A..28F}
{Fouesneau} M.,  et~al., 2023, \mn@doi [\aap] {10.1051/0004-6361/202243919},
  \href {https://ui.adsabs.harvard.edu/abs/2023A&A...674A..28F} {674, A28}

\bibitem[\protect\citeauthoryear{{Freudenthal} et~al.,}{{Freudenthal}
  et~al.}{2019}]{Freudenthal19}
{Freudenthal} J.,  et~al., 2019, \mn@doi [\aap] {10.1051/0004-6361/201935879},
  \href {https://ui.adsabs.harvard.edu/abs/2019A&A...628A.108F} {628, A108}

\bibitem[\protect\citeauthoryear{Fulton \& Petigura}{Fulton \&
  Petigura}{2018}]{Fulton2018}
Fulton B.~J.,  Petigura E.~A.,  2018, \mn@doi [\aj] {10.3847/1538-3881/aae828},
  156, 264

\bibitem[\protect\citeauthoryear{{Gaia Collaboration} et~al.,}{{Gaia
  Collaboration} et~al.}{2018}]{2018A&A...616A...1G}
{Gaia Collaboration} et~al., 2018, \mn@doi [\aap]
  {10.1051/0004-6361/201833051}, \href
  {https://ui.adsabs.harvard.edu/abs/2018A&A...616A...1G} {616, A1}

\bibitem[\protect\citeauthoryear{{Gelman} \& {Rubin}}{{Gelman} \&
  {Rubin}}{1992}]{GelmanRubin92}
{Gelman} A.,  {Rubin} D.~B.,  1992, \mn@doi [Statistical Science]
  {10.1214/ss/1177011136}, \href
  {https://ui.adsabs.harvard.edu/abs/1992StaSc...7..457G} {7, 457}

\bibitem[\protect\citeauthoryear{Hadden \& Lithwick}{Hadden \&
  Lithwick}{2014}]{HL14}
Hadden S.,  Lithwick Y.,  2014, \mn@doi [\apj] {10.1088/0004-637x/787/1/80},
  787, 80

\bibitem[\protect\citeauthoryear{Hadden \& Lithwick}{Hadden \&
  Lithwick}{2016}]{HL16}
Hadden S.,  Lithwick Y.,  2016, \mn@doi [\apj] {10.3847/0004-637x/828/1/44},
  828, 44

\bibitem[\protect\citeauthoryear{Hadden \& Lithwick}{Hadden \&
  Lithwick}{2017}]{HL17}
Hadden S.,  Lithwick Y.,  2017, \mn@doi [\aj] {10.3847/1538-3881/aa71ef}, 154,
  5

\bibitem[\protect\citeauthoryear{Holczer et~al.,}{Holczer
  et~al.}{2016}]{Holczer16}
Holczer T.,  et~al., 2016, \mn@doi [{\apj} Supplement Series]
  {10.3847/0067-0049/225/1/9}, 225, 9

\bibitem[\protect\citeauthoryear{Holman}{Holman}{2005}]{Holman05}
Holman M.~J.,  2005, \mn@doi [Science] {10.1126/science.1107822}, 307, 1288

\bibitem[\protect\citeauthoryear{Jontof-Hutter et~al.,}{Jontof-Hutter
  et~al.}{2016}]{JH16}
Jontof-Hutter D.,  et~al., 2016, \mn@doi [\apj] {10.3847/0004-637x/820/1/39},
  820, 39

\bibitem[\protect\citeauthoryear{{Jontof-Hutter}, {Wolfgang}, {Ford},
  {Lissauer}, {Fabrycky}  \& {Rowe}}{{Jontof-Hutter} et~al.}{2021}]{JH21}
{Jontof-Hutter} D.,  {Wolfgang} A.,  {Ford} E.~B.,  {Lissauer} J.~J.,
  {Fabrycky} D.~C.,   {Rowe} J.~F.,  2021, \mn@doi [\aj]
  {10.3847/1538-3881/abd93f}, \href
  {https://ui.adsabs.harvard.edu/abs/2021AJ....161..246J} {161, 246}

\bibitem[\protect\citeauthoryear{{Jontof-Hutter}, {Dalba}  \&
  {Livingston}}{{Jontof-Hutter} et~al.}{2022}]{JH22}
{Jontof-Hutter} D.,  {Dalba} P.~A.,   {Livingston} J.~H.,  2022, \mn@doi [\aj]
  {10.3847/1538-3881/ac7396}, \href
  {https://ui.adsabs.harvard.edu/abs/2022AJ....164...42J} {164, 42}

\bibitem[\protect\citeauthoryear{{Judkovsky}, {Ofir}  \&
  {Aharonson}}{{Judkovsky} et~al.}{2021a}]{Judkovsky22a}
{Judkovsky} Y.,  {Ofir} A.,   {Aharonson} O.,  2021a, arXiv e-prints, \href
  {https://ui.adsabs.harvard.edu/abs/2021arXiv211210132J} {p. arXiv:2112.10132}

\bibitem[\protect\citeauthoryear{{Judkovsky}, {Ofir}  \&
  {Aharonson}}{{Judkovsky} et~al.}{2021b}]{Judkovsky22b}
{Judkovsky} Y.,  {Ofir} A.,   {Aharonson} O.,  2021b, arXiv e-prints, \href
  {https://ui.adsabs.harvard.edu/abs/2021arXiv211210144J} {p. arXiv:2112.10144}

\bibitem[\protect\citeauthoryear{{Judkovsky}, {Ofir}  \&
  {Aharonson}}{{Judkovsky} et~al.}{2023}]{Judkovsky23}
{Judkovsky} Y.,  {Ofir} A.,   {Aharonson} O.,  2023, \mn@doi [\aj]
  {10.3847/1538-3881/ad07ce}, \href
  {https://ui.adsabs.harvard.edu/abs/2023AJ....166..256J} {166, 256}

\bibitem[\protect\citeauthoryear{{Latham}, {Brown}, {Monet}, {Everett},
  {Esquerdo}  \& {Hergenrother}}{{Latham} et~al.}{2005}]{Latham05}
{Latham} D.~W.,  {Brown} T.~M.,  {Monet} D.~G.,  {Everett} M.,  {Esquerdo}
  G.~A.,   {Hergenrother} C.~W.,  2005, in American Astronomical Society
  Meeting Abstracts. p. 110.13

\bibitem[\protect\citeauthoryear{{Libby-Roberts} et~al.,}{{Libby-Roberts}
  et~al.}{2020}]{LibbyRoberts20}
{Libby-Roberts} J.~E.,  et~al., 2020, \mn@doi [\aj] {10.3847/1538-3881/ab5d36},
  \href {https://ui.adsabs.harvard.edu/abs/2020AJ....159...57L} {159, 57}

\bibitem[\protect\citeauthoryear{{Lissauer} et~al.,}{{Lissauer}
  et~al.}{2013}]{Lissauer13}
{Lissauer} J.~J.,  et~al., 2013, \mn@doi [\apj] {10.1088/0004-637X/770/2/131},
  \href {https://ui.adsabs.harvard.edu/abs/2013ApJ...770..131L} {770, 131}

\bibitem[\protect\citeauthoryear{Lithwick, Xie  \& Wu}{Lithwick
  et~al.}{2012}]{LithXieWu12}
Lithwick Y.,  Xie J.,   Wu Y.,  2012, \mn@doi [\apj]
  {10.1088/0004-637x/761/2/122}, 761, 122

\bibitem[\protect\citeauthoryear{{MacDonald} et~al.,}{{MacDonald}
  et~al.}{2016}]{MacDonald16}
{MacDonald} M.~G.,  et~al., 2016, \mn@doi [\aj] {10.3847/0004-6256/152/4/105},
  \href {https://ui.adsabs.harvard.edu/abs/2016AJ....152..105M} {152, 105}

\bibitem[\protect\citeauthoryear{Mandel \& Agol}{Mandel \& Agol}{2002}]{MA02}
Mandel K.,  Agol E.,  2002, \mn@doi [\apj] {10.1086/345520}, 580, L171

\bibitem[\protect\citeauthoryear{{Masuda}, {Hirano}, {Taruya}, {Nagasawa}  \&
  {Suto}}{{Masuda} et~al.}{2013}]{masuda13}
{Masuda} K.,  {Hirano} T.,  {Taruya} A.,  {Nagasawa} M.,   {Suto} Y.,  2013,
  \mn@doi [\apj] {10.1088/0004-637X/778/2/185}, \href
  {https://ui.adsabs.harvard.edu/abs/2013ApJ...778..185M} {778, 185}

\bibitem[\protect\citeauthoryear{Murray \& Dermott}{Murray \&
  Dermott}{2000}]{Murray2000}
Murray C.~D.,  Dermott S.~F.,  2000, Solar System Dynamics.
Cambridge University Press, \mn@doi{10.1017/cbo9781139174817}, \url
  {https://doi.org/10.1017/cbo9781139174817}

\bibitem[\protect\citeauthoryear{{NASA Exoplanet Archive}}{{NASA Exoplanet
  Archive}}{2022}]{koidr25}
{NASA Exoplanet Archive} 2022, Kepler Objects of Interest DR25,
  \mn@doi{10.26133/NEA5}, \url
  {https://catcopy.ipac.caltech.edu/dois/doi.php?id=10.26133/NEA19}

\bibitem[\protect\citeauthoryear{Ofir \& Dreizler}{Ofir \&
  Dreizler}{2013}]{Ofir2013}
Ofir A.,  Dreizler S.,  2013, \mn@doi [\aap] {10.1051/0004-6361/201219877},
  555, A58

\bibitem[\protect\citeauthoryear{{Ofir}, {Xie}, {Jiang}, {Sari}  \&
  {Aharonson}}{{Ofir} et~al.}{2018}]{Ofir18}
{Ofir} A.,  {Xie} J.-W.,  {Jiang} C.-F.,  {Sari} R.,   {Aharonson} O.,  2018,
  \mn@doi [\apjs] {10.3847/1538-4365/aa9f2b}, \href
  {http://adsabs.harvard.edu/abs/2018ApJS..234....9O} {234, 9}

\bibitem[\protect\citeauthoryear{Pastore \& Calcagnì}{Pastore \&
  Calcagnì}{2019}]{Pastore2019}
Pastore M.,  Calcagnì A.,  2019, \mn@doi [Frontiers in Psychology]
  {10.3389/fpsyg.2019.01089}, 10, 1089

\bibitem[\protect\citeauthoryear{{Santerne} et~al.,}{{Santerne}
  et~al.}{2016}]{Santerne2016}
{Santerne} A.,  et~al., 2016, \mn@doi [\aap] {10.1051/0004-6361/201527329},
  \href {https://ui.adsabs.harvard.edu/abs/2016A&A...587A..64S} {587, A64}

\bibitem[\protect\citeauthoryear{{Schmitt} et~al.,}{{Schmitt}
  et~al.}{2014}]{Schmitt2014}
{Schmitt} J.~R.,  et~al., 2014, \mn@doi [\apj] {10.1088/0004-637X/795/2/167},
  \href {https://ui.adsabs.harvard.edu/abs/2014ApJ...795..167S} {795, 167}

\bibitem[\protect\citeauthoryear{{Shallue} \& {Vanderburg}}{{Shallue} \&
  {Vanderburg}}{2018}]{Shallue18}
{Shallue} C.~J.,  {Vanderburg} A.,  2018, \mn@doi [\aj]
  {10.3847/1538-3881/aa9e09}, \href
  {https://ui.adsabs.harvard.edu/abs/2018AJ....155...94S} {155, 94}

\bibitem[\protect\citeauthoryear{{Vissapragada} et~al.,}{{Vissapragada}
  et~al.}{2020}]{Vissapragada20}
{Vissapragada} S.,  et~al., 2020, \mn@doi [\aj] {10.3847/1538-3881/ab65c8},
  \href {https://ui.adsabs.harvard.edu/abs/2020AJ....159..108V} {159, 108}

\bibitem[\protect\citeauthoryear{{Weiss} et~al.,}{{Weiss}
  et~al.}{2013}]{Weiss13}
{Weiss} L.~M.,  et~al., 2013, \mn@doi [\apj] {10.1088/0004-637X/768/1/14},
  \href {https://ui.adsabs.harvard.edu/abs/2013ApJ...768...14W} {768, 14}

\bibitem[\protect\citeauthoryear{Xie}{Xie}{2013}]{Xie13}
Xie J.-W.,  2013, \mn@doi [{\apj} Supplement Series]
  {10.1088/0067-0049/208/2/22}, 208, 22

\bibitem[\protect\citeauthoryear{Xie}{Xie}{2014}]{Xie14}
Xie J.-W.,  2014, \mn@doi [{\apj} Supplement Series]
  {10.1088/0067-0049/210/2/25}, 210, 25

\bibitem[\protect\citeauthoryear{{Yoffe}, {Ofir}  \& {Aharonson}}{{Yoffe}
  et~al.}{2021}]{Yoffe21}
{Yoffe} G.,  {Ofir} A.,   {Aharonson} O.,  2021, \mn@doi [\apj]
  {10.3847/1538-4357/abc87a}, \href
  {https://ui.adsabs.harvard.edu/abs/2021ApJ...908..114Y} {908, 114}

\makeatother
\end{thebibliography}


\label{lastpage}
\end{document}